\documentclass[twocolumn,superscriptaddress,prl]{revtex4-2}

\usepackage[utf8]{inputenc}
\usepackage{graphicx}
\usepackage{tikz}
\usepackage{xspace}
\usetikzlibrary{calc}
\usepackage{booktabs}
\usepackage{makecell}
\usetikzlibrary{positioning,fit}

\usepackage[colorlinks=true, allcolors=black]{hyperref}
\usepackage[capitalize,noabbrev]{cleveref}
\usepackage{placeins}

\crefname{section}{Section}{Sections}
\crefname{appendix}{Appendix}{Appendices}
\Crefname{appendix}{Appendix}{Appendices}

\newcommand{\device}{\texttt{ibm\_phoenix}\xspace}
\newcommand{\ket}[1]{\left|#1\right\rangle}

\begin{document}

\title{Entangled logical qubits on a superconducting quantum processor}

\author{Robin Harper}
\affiliation{School of Physics, The University of Sydney, Sydney, New South Wales 2006, Australia}

\author{Xanda C. Kolesnikow}
\affiliation{School of Physics, The University of Sydney, Sydney, New South Wales 2006, Australia}

\author{Campbell K. McLauchlan}
\affiliation{School of Physics, The University of Sydney, Sydney, New South Wales 2006, Australia}

\author{Georgia M. Nixon}
\affiliation{School of Physics, The University of Sydney, Sydney, New South Wales 2006, Australia}

\author{Stephen D. Bartlett}
\email{stephen.bartlett@sydney.edu.au}
\affiliation{School of Physics, The University of Sydney, Sydney, New South Wales 2006, Australia}

\author{Benjamin J. Brown}
\affiliation{IBM Quantum, T. J. Watson Research Center, Yorktown Heights, New York 10598, USA}
\affiliation{IBM Denmark, Sundkrogsgade 11, 2100 Copenhagen, Denmark}

\begin{abstract}
We demonstrate high-fidelity logical qubit operations using \texttt{ibm\_phoenix}, a square-lattice array of 120 coupled superconducting qubits that are capable of fast reset. Using strategies to exclude underperforming components together with noise informed decoding, we routinely produce surface codes at distance-3 and distance-5 with logical error per round below 1\%.  We demonstrate entanglement between two logical qubits, where logical Bell-state fidelities above 93\% are observed for entanglement created via lattice surgery.  With three logical qubits and lattice surgery, we perform logical teleportation of a range of states and gates, observing fidelities in the range 78\% - 93\%. Using reset to reduce qubit leakage, we show that repeated rounds of stabilizer measurements continue to exhibit improvements in a stability experiment up to $\sim 2\cdot 10^{-7}$ over 170 measurement rounds. 
\end{abstract}

\maketitle

A quantum computer that can execute large, useful quantum circuits will require near-noiseless logical qubits capable of performing a universal set of logic gates. A pathway to building such a computer at scale is to use many coupled physical qubits in an error-corrected computing architecture. 
As we continue to develop increasingly sophisticated quantum processors, it is important that we continually test and evaluate their capability of performing computational tasks while simultaneously identifying and reversing the impact of errors. 
Indeed, for any hardware platform, schemes for quantum error correction must be designed to account for the typical errors that the quantum processor will experience in order to optimize performance. 

In the case of superconducting qubits, significant recent work has explored quantum error correction in experiment~\cite{Ye_magicstateprep_2023,Gupta2024,google24,lacroix2024scalinglogiccolorcode,caune_demonstrating_2024,bence2024,harper2025characterising,google_magic2025,besedin_lattice_2026,lee2026scalable,wang2026super,drouet2026defects, lin2026surface, mclauchlan2026,lange2026,kan2026demonstrationlogicalerrorsuppression}. In addition to errors acting on the qubit subspace, two additional barriers to high-fidelity logical qubits have been identified.  
The first is leakage out of the qubit subspace~\cite{Aliferis2007, Fowler_2013, Suchara_2015}, which if left uncorrected can compound throughout a deep circuit~\cite{Varbanov2020,mcewen2021,miao_overcoming_2023,harper2025characterising,lee2026scalable,mclauchlan2026}. 
The second is apparent on large devices with hundreds of components, where some fraction of these components underperform due to variability in the fabrication process.  These underperforming components can have a significant effect on logical performance~\cite{Stace2009thresholds, Auger2017, Strikis2021, Siegel2023adaptivesurfacecode, Lin2024codesign, McLauchlan2024accommodating, Wei2025low, Leroux2025snakes, Debroy2025luciinsurfacecode, mishmash2025excising, wolanski2026automated, wei2026adaptivedeformationcolorcode,higgott2025handlingfabricationdefectshexgrid, drouet2026defects, mclauchlan2026}. 
If left unaddressed, both of these processes are detrimental to logical performance because they are not corrected by standard error-correcting codes.

Here we demonstrate quantum memory, entangled logical qubits and logic operations on the IBM Nighthawk revision 2 quantum processor \texttt{ibm\_phoenix}. 
We demonstrate surface-code quantum memories~\cite{Kitaev1997a, Dennis2002, Fowler2012} with a logical error per round below $1\%$ using both distance-3 and distance-5 surface codes. 
To accommodate variability in the physical qubit performance, we make essential use of defect exclusion techniques~\cite{Stace2009thresholds, Auger2017, Strikis2021, Siegel2023adaptivesurfacecode, Lin2024codesign, McLauchlan2024accommodating, Wei2025low, Leroux2025snakes, Debroy2025luciinsurfacecode, mishmash2025excising, wolanski2026automated, wei2026adaptivedeformationcolorcode,higgott2025handlingfabricationdefectshexgrid, drouet2026defects, mclauchlan2026} to avoid underperforming qubits.  
These techniques halve the logical error per round in a distance-5 memory experiment, from $1.9\%$ to $0.9\%$. We also demonstrate logical entanglement between pairs of logical qubits using lattice surgery~\cite{Horsman2012}, with logical Bell-state fidelity above $93\%$ using defect exclusion.  
Finally, using three logical qubits, we demonstrate both state and gate teleportation.  
This logical performance is facilitated by a number of key innovations that address the dominant error processes described above.  
The quantum processor \texttt{ibm\_phoenix} possesses a fast reset operation on qubits, which we leverage to improve quantum error correction performance and to reduce leakage error processes in our experiments, with considerable effect.  
As a striking illustration of the effect of this new capability, we present a stability experiment~\cite{Gidney2022dual} on a $4\times 4$ square patch of the device over $170$ rounds of measurement that achieves overall probability of error approaching $10^{-7}$.

\paragraph{High-fidelity logical qubits.}
\label{sec:threshold}

\begin{figure*}
    \centering
    \centering
\begin{tikzpicture}[
    inner sep=5pt,
    outer sep=0pt
]

\node (round1)  at (-6.8,0) {
    \includegraphics[width=0.34\linewidth]{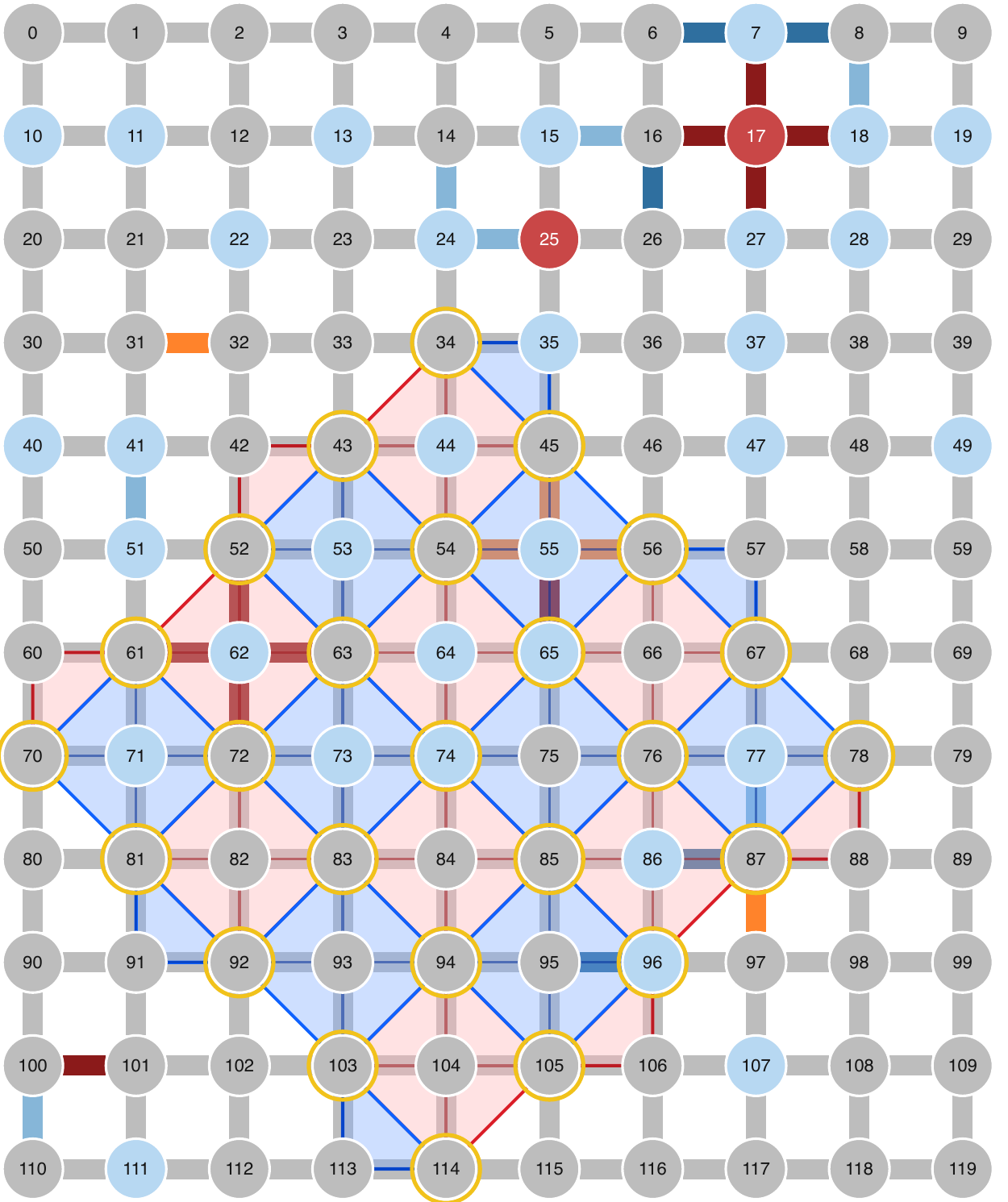}
};

\node (round0) at (2.8,0) {
        \includegraphics[width=0.6\linewidth]{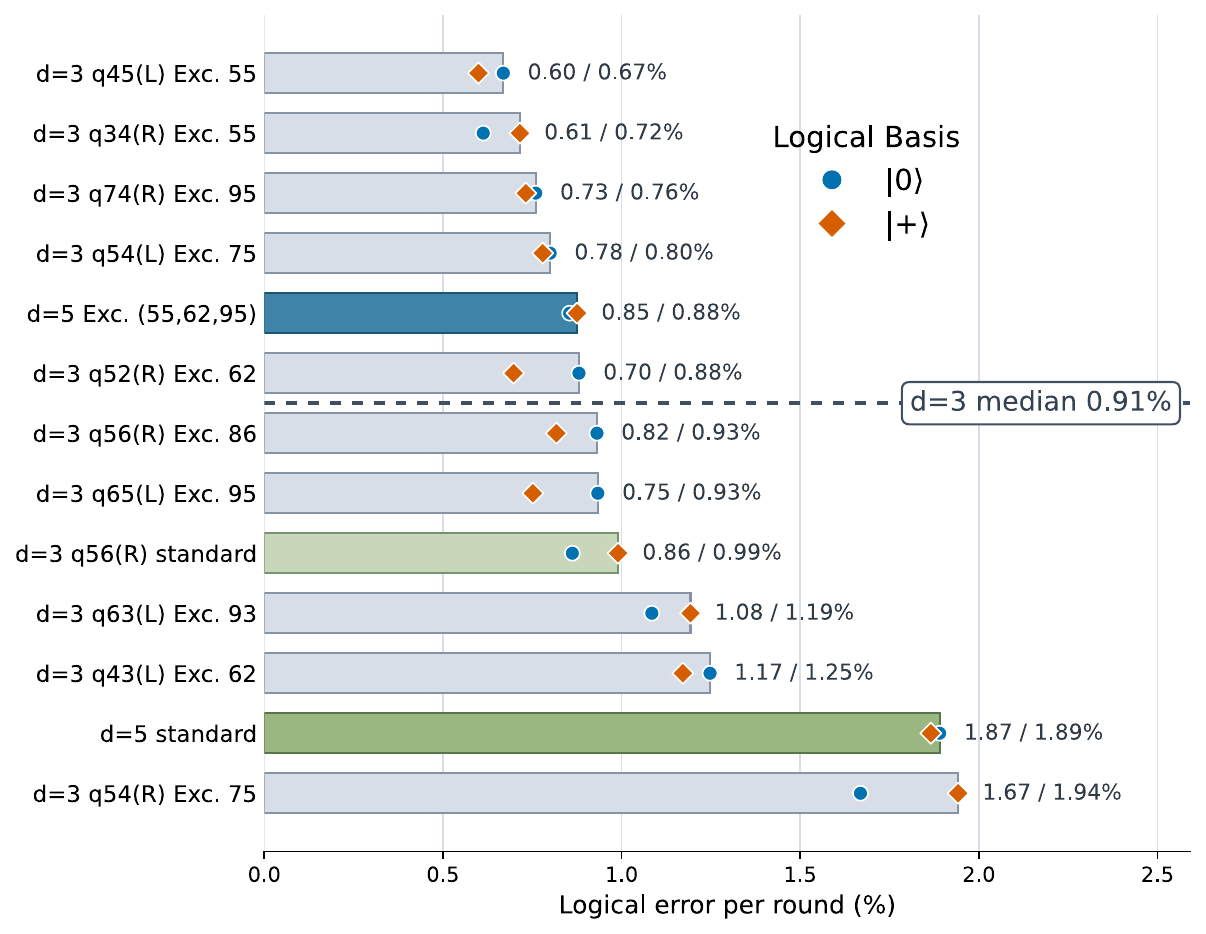}
};

\node (round1)  at (-6.9,-4) {
    \includegraphics[width=0.32\linewidth]{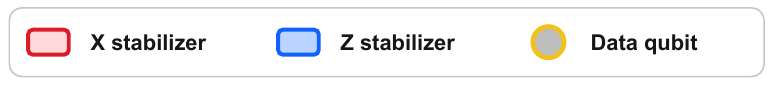}
};
\node (round1)  at (-3.8,1.9) {
    \includegraphics[width=0.14\linewidth]{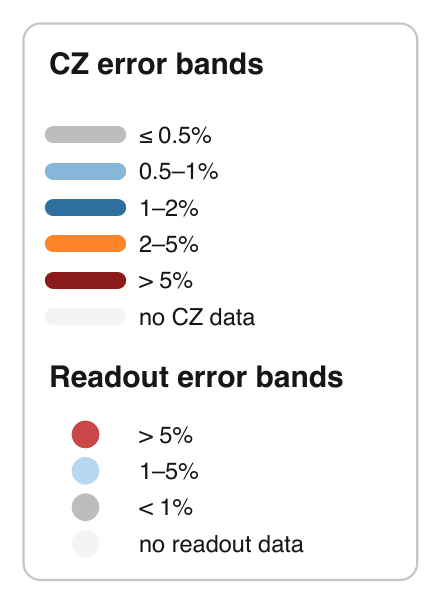}
};
\end{tikzpicture}
    
        \caption{\textbf{Memory experiments.} (\textbf{Left}) Placement of a distance-5 surface code on the quantum processor \device. Codes are labeled by qubit number using the top qubit (lowest index) of the code, and a direction indicator (L for left or R  for right) indicating where the two-body stabilizer lies in relation to the top qubit of the code. For the layout shown this code is q34(R). (\textbf{Right}) Logical error per round for memory experiments with distance-5 and distance-3 surface codes. ``Exc.''~(excluded) indicates the ancilla qubits that we re-purposed, details can be found in Supplemental Material S2; ``standard'' indicates the unmodified surface code syndrome extraction circuit with no ancilla re-purposing, for reference (these are highlighted with green bars).  All distance-3 codes shown lie in the footprint of the distance-5 code. The median logical error per round for the best distance-3 codes in each location is shown (dashed line). Distance-3 (distance-5) codes with ancilla re-purposing have bars colored grey (blue). 
        }
    \label{fig:identity}
\end{figure*}

Quantum error correction provides a pathway for near-noiseless logical operations on a quantum computer. 
Using a quantum processor with sufficiently high-fidelity components, we can construct codes that are capable of detecting and reversing the effect of errors, thereby suppressing the probability of logical errors. 
By producing larger codes with a larger code distance capable of correcting a larger number of errors, we expect that we can suppress logical errors arbitrarily.

A major challenge to achieving high-fidelity logical qubits on IBM Nighthawk quantum processors and similar devices is the variability in component performance~\cite{drouet2026defects, mclauchlan2026}. 
While theory and simulations predict that the average performance of the components of \device are below an error threshold, where increasing the code distance should suppress logical errors, we find that underperforming qubits limit this logical error suppression significantly. 
This is evident from the improvements in logical error per round in memory experiments that we observe using defect exclusion strategies to avoid the use of these underperforming components. 
Naturally, for a given quantum processor with some configuration of defective components, the impact of underperforming components is more marked on larger codes due to the fact that the placement of a larger code is more likely to collide with a greater number of defective parts. 

Here, we demonstrate that with defect exclusion techniques, the performance of larger distance codes with distance-$5$ is comparable with the performance of the highest-performing distance-3 codes.  
\cref{fig:identity} presents the results of memory experiments using a distance-5 code together with the spectrum of distance-3 codes that sit within the larger code's footprint.  
We use ancilla re-purposing to selectively remove up to three of the qubits in the distance-5 code (see Supplemental Material S1 for details).  
For each distance-3 code, we experimentally assess all 4 single-ancilla re-purposing strategies, always identifying at least one ancilla where re-purposing provides an advantage (see Supplemental Material S2).  
The logical error per round of the distance-5 code using ancilla re-purposing is lower than the median distance-3 code in the larger code's footprint, also using ancilla re-purposing.  
Decoding is performed with a Belief Matching~\cite{higgott2023improved} decoder that is supplied with a detailed noise model (see Supplemental Material S3); the need for ancilla re-purposing to achieve logical errors per round below 1\% demonstrates that informing the decoder about the underperforming components is not itself sufficient and that we must actively adapt the syndrome extraction circuits. This was also observed in Ref.~\cite{drouet2026defects}.

\paragraph{Entangled logical qubits.}
\label{sec:AandB}

\begin{figure}
\begin{tikzpicture}

\node  at (-3,-0.1) {
    \includegraphics[width=0.52\linewidth]{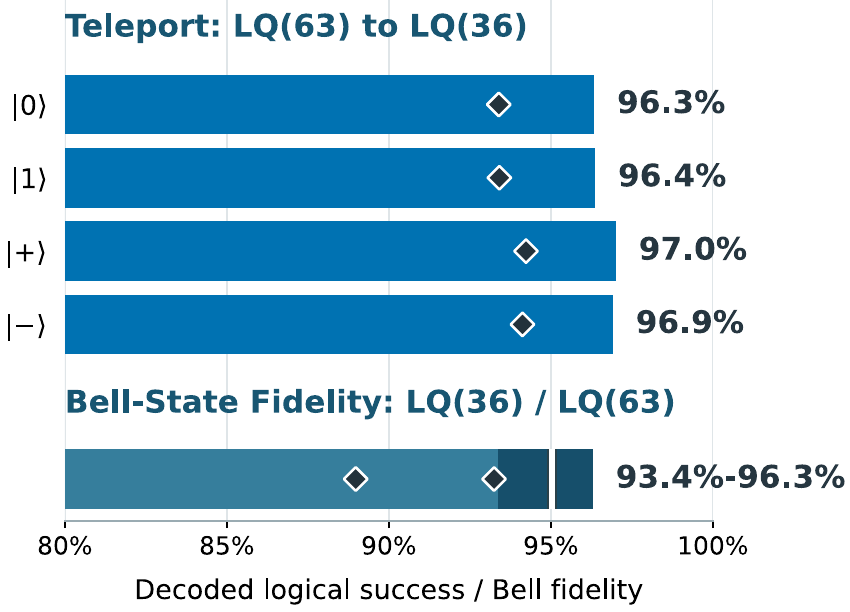}
};
\node  at (1.1,-0.1) {
    \includegraphics[width=0.5\linewidth]{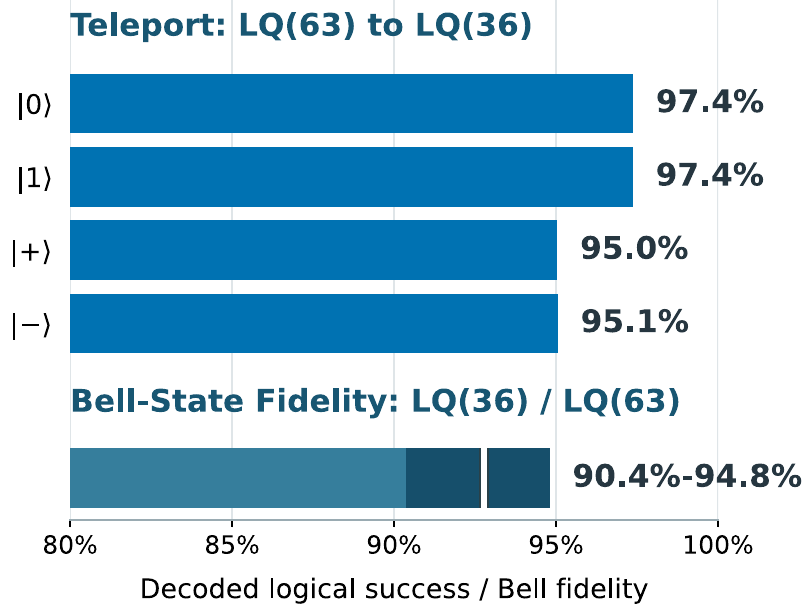}
};
\node  at (-5.3,1.45) {
    (a)
};
\node  at (-1.1,1.45) {
    (b)
};

\end{tikzpicture}

    \caption{\textbf{Entangled logical qubits.} (\textbf{a}) Here we show the results of using lattice surgery (in this case an XX measurement) to (i) teleport cardinal states from one logical qubit to the other (top); and (ii) create a Bell state between the two logical qubits (bottom). Placement of the code and details of the schedule used can be found in Supplemental Material S4. The experiments were run on 6 October, 2026, with 500,000 shots taken per circuit. Only shots where the logical qubits are initialized in the code-state prior to lattice surgery are used, by post-selecting out any non-trivial detectors for the code preparation rounds; the yield in each case is $40\%\pm2\%$.  The diamond shows the results in the case where no code-space post-selection is applied. The Bell-state fidelity is determined using only the $\langle$XX$\rangle$ and $\langle$ZZ$\rangle$ expectation values (we do not estimate $\langle$YY$\rangle$) to provide upper and lower bounds of fidelity of the Bell state. The white line indicates the value of the Bell-state fidelity with a depolarizing noise assumption. The diamonds indicate the upper and lower bounds of the Bell-state fidelity without any post-selection.\textbf{(b)} Experiments run on 9 November (teleport) and 29 October (Bell-state fidelity). Here no memory circuits were run prior to entangling the logical qubits. No post-selection was performed. The $\langle$XX$\rangle+\langle$ZZ$\rangle$ expectation value is 1.81.}
    \label{fig:two_results}
\end{figure}

A key primitive in fault-tolerant quantum computing is lattice surgery, which provides a fault-tolerant implementation of a logical entangling gate.  
In lattice surgery, two adjacent surface codes are combined by measuring local operators in the intermediate region between codes (which we call the seam), thereby entangling the logical qubits using a joint logical ZZ or XX measurement (depending on the edges that were combined), followed by splitting the codes and performing error correction to return to the original code space. 
For example, if both codes are prepared in the logical $|0\rangle$ state and a logical XX measurement is performed, then, depending on the measurement outcome, they will be projected into one of the logical Bell states $\frac{1}{\sqrt{2}}(|00\rangle\pm|11\rangle)$.  
This lattice surgery operation can be used to deterministically prepare a logical Bell state.  
In addition, we can demonstrate one-qubit teleportation~\cite{PERDRIX_2005}, also known as state transfer, from one logical qubit to the other.

We present results for our experiment demonstrating entangled logical qubits in \Cref{fig:two_results}. 
To investigate the performance specifically of the entangling (lattice surgery) operation, we analyse only those shots where we have confidently initialized in the code space prior to lattice surgery.  
For this, we check for trivial detector activity once the logical qubits have been projected into the code space and only proceed if this is the case. (Alternatively, one could use prepare-test-repeat circuits to initialize into the code space~\cite{Gupta2024}.) 

We report the fidelity of the logical Bell state created in~\Cref{fig:two_results}. 
Because of the high SPAM error on the Y readout (see Supplemental Material S1.2), we only measure the logical XX and ZZ observables, and provide bounds on the logical Bell-state fidelity using  the $\langle$ZZ$\rangle$  and $\langle$XX$\rangle$  expectation values; see Ref.~\cite{mclauchlan2026}. 
For the teleportation experiment we prepare the first logical qubit in one of the four cardinal states and after teleportation measure the recipient logical qubit in the corresponding basis. 
The results reported are the percentage of successful measurements. Supplemental Material S4 presents additional details.

\begin{figure}
    \begin{tikzpicture}[
        inner sep=5pt,
        outer sep=0pt
    ]

    \node  at (0,0) {
        \includegraphics[width=1\linewidth]{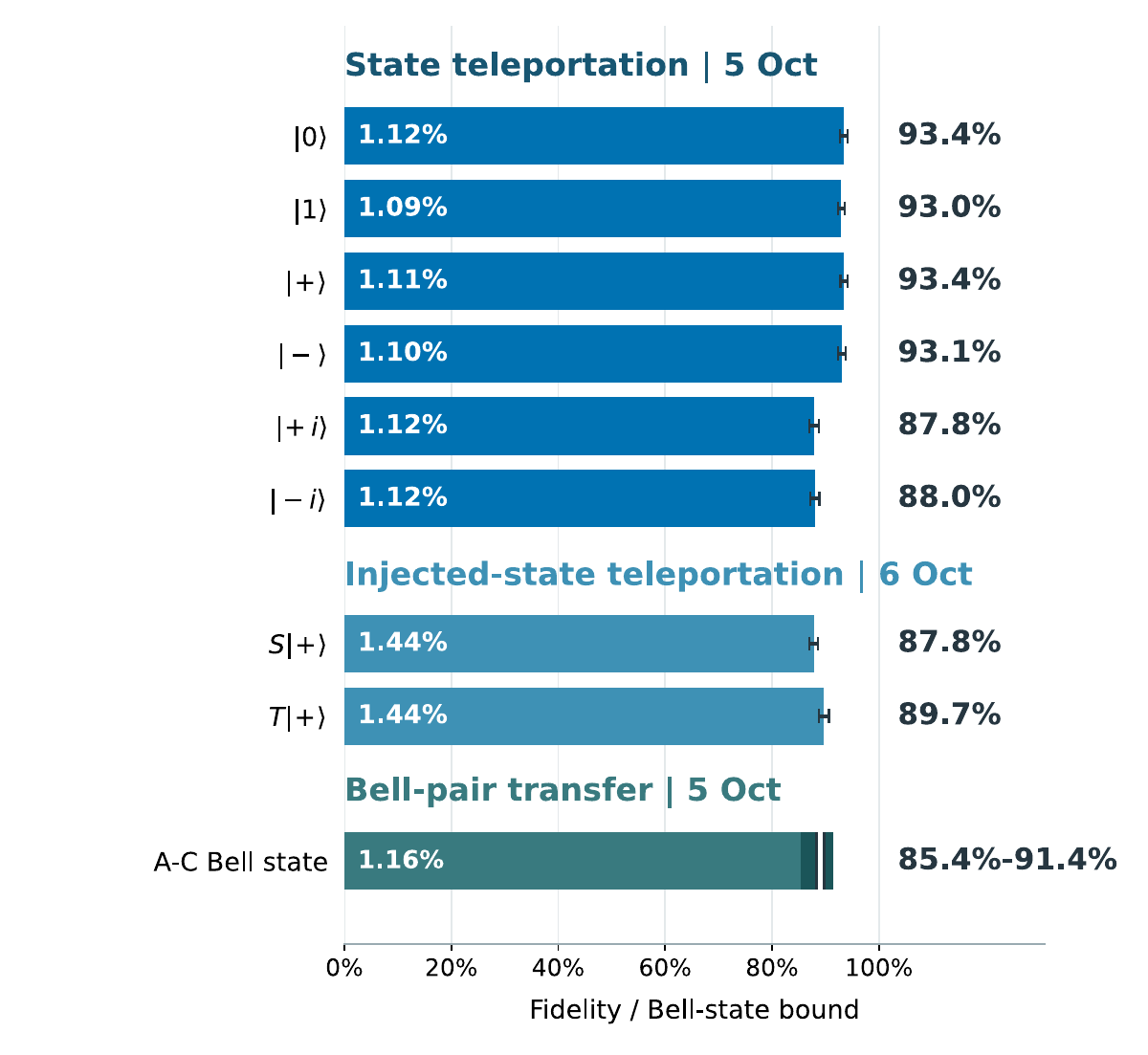}
    };

    \node at (0,-10) {
            \includegraphics[width=1\linewidth]{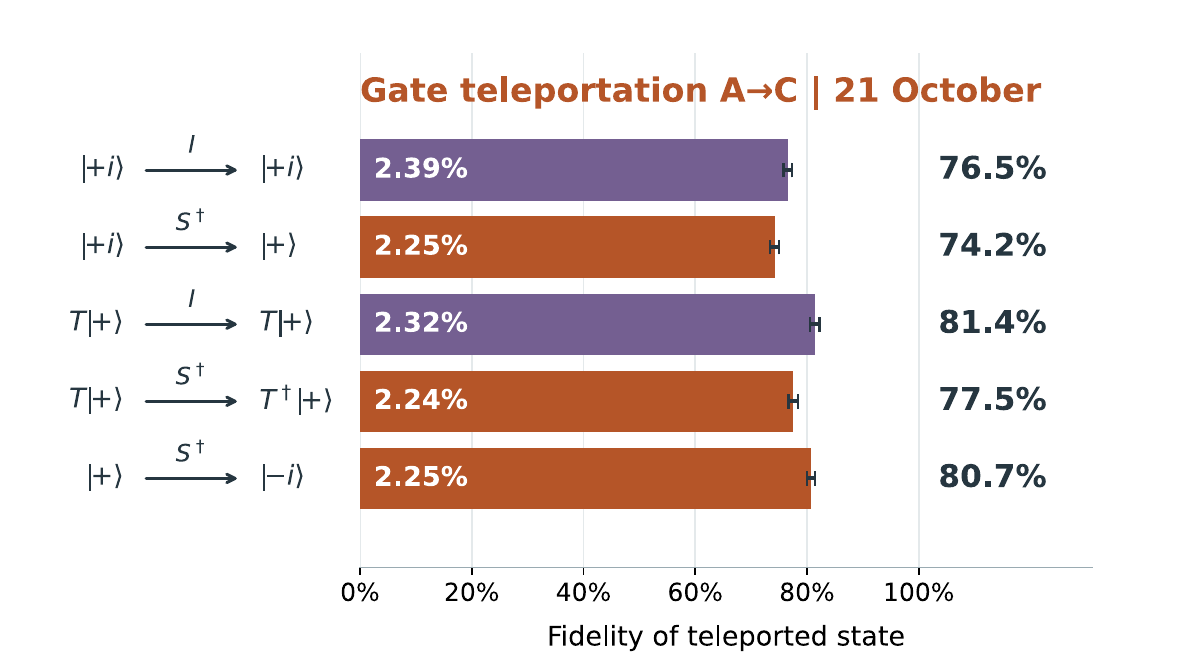}
    };
    \node  at (0,-5.8) {
            \includegraphics[width=1\linewidth]{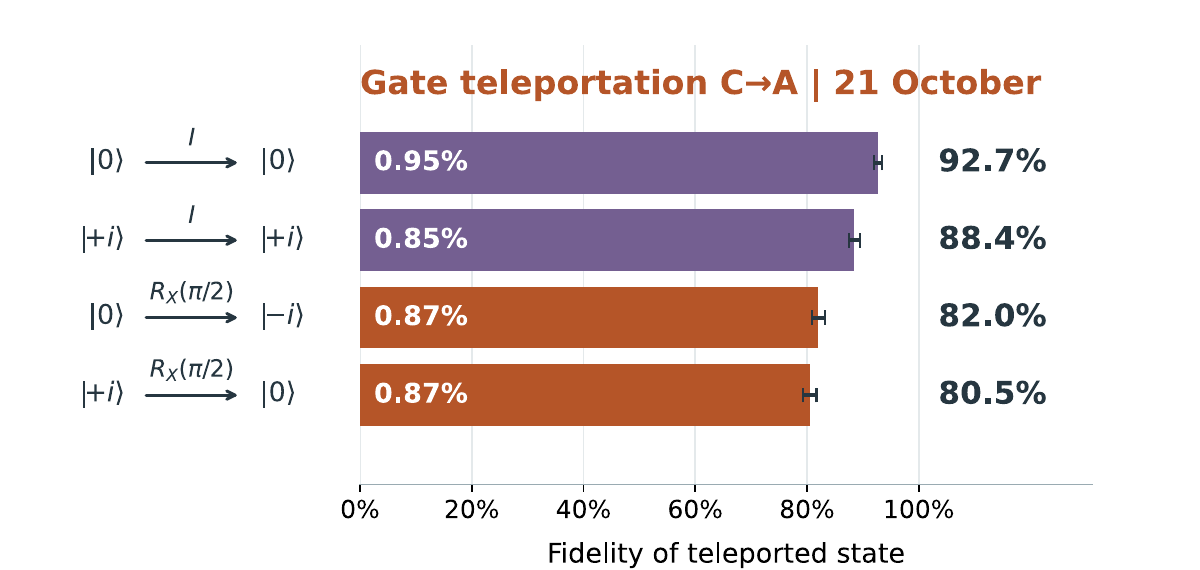}
    };

    \node at (-3,4) {(a)};
    \node at (-3,-4) {(b)};
    \node at (-3,-8.2) {(c)};

    \end{tikzpicture}
    \caption{\textbf{Teleporting logical qubits.}  \textbf{(a)} Logical fidelities for state teleportation, injected-state teleportation, and Bell-pair transfer. All are single-logical-qubit fidelities, with the exception of the Bell state fidelity where expectation values of logical XX and ZZ provide a bound. \textbf{(b)--(c)} Logical fidelities for gate teleportation experiments, either from Alice to Charlie \textbf{(b)} or Charlie to Alice \textbf{(c)}. For all demonstrations in \textbf{(a)--(c)}, we collect 500,000 shots. White text on bars indicates yields from post-selection on the correct initial state preparation. Details of circuit placement can be found in Supplemental Material S5.}
    \label{fig:threeResults}
\end{figure}

Performing experiments involving three logical qubits placed on a 120-qubit IBM Nighthawk processor requires care in the placement of the codes, as well as how lattice surgery is executed.  
Typically, lattice surgery proposals require a region of qubits between the two logical qubit code regions~\cite{Horsman2012,fowler2012b,Litinski2019gameofsurfacecodes,fowler2019lowoverheadquantumcomputation,bodeker2026,lin2026surfacecodelogicaloperations}. Ref.~\cite{bodeker2026} noted that the logical error rate increases as the width of the region between the two logical qubits is increased, and earlier experiments on other IBM quantum processors~\cite{harper2025characterising,drouet2026defects,lin2026surfacecodelogicaloperations} indicated that surgery between adjacent logical qubits would be optimal. 

A region of qubits between the codes to implement lattice surgery is typically proposed because otherwise, the two logical qubits compete for the use of ancilla along the seam.  We overcome this requirement by either allowing the codes to use the ancilla on alternate rounds (thereby completing a detector with the weight-two stabilizer half as frequently) or re-purposing the ancilla of one of the logical qubits (see Supplemental Material S4 and S5). The re-purposing causes the two-body stabilizer to be measured on the ancilla of the adjoining four-body stabilizer on alternate rounds. Since it commutes with all the other stabilizers, no gauge detectors need be formed. In this case only the logical qubit with the re-purposed ancilla has changed detectors. Both the edge two-body detector and the four-body detector with the re-purposed ancilla span two rounds, rather than the usual one round.

Lattice surgery proceeds as normal, save that  during lattice surgery the combined codes form a $3 \times 6$ logical qubit, rather than the traditional $3\times 7$ size.

\paragraph{Teleporting logical qubits.}
\label{sec:three}

\begin{figure}
    \centering
       \begin{tikzpicture}

  \node[anchor=south west, inner sep=0] (img) at (-0.5,0) {%
    \includegraphics[width=1\linewidth]{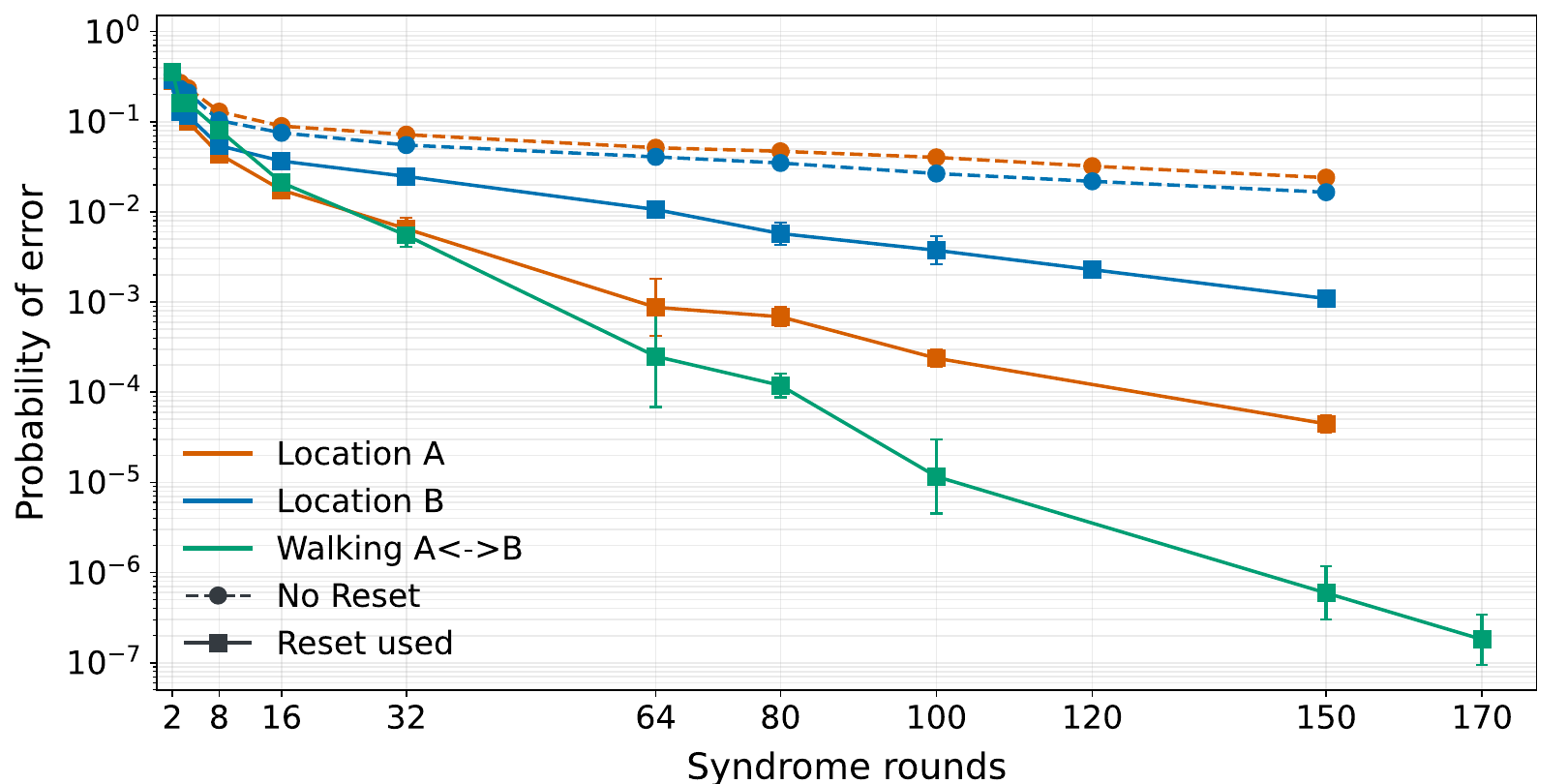}};

    \end{tikzpicture}

    \caption{\textbf{Stability experiments with fast reset.} Error rate after repeated rounds of the stability experiment. Unlike previous devices, \device has an unconditional reset.  We compare stability experiments where ancillas are reset after every stabilizer measurement, against stability experiments where no reset is used. }
    \label{fig:stability}
\end{figure}

By compressing seams in lattice surgery, there is sufficient space on \device to place a third logical qubit in a position that permits lattice surgery between the middle logical qubit and the other two, albeit along different boundary types. Supplemental Material S5 provides details on how these logical qubits are entangled.  The logical three-qubit experiments described below all initialize a Bell state on two of the logical qubits and a cardinal state on the third. 
We post-select on these initializations, rejecting those runs where we detect errors while preparing the initial Bell state or projecting the third logical qubit into the correct state.

With the supplied logical Bell pair between Alice and Bob, and Charlie possessing an arbitrary state encoded in her logical qubit, we can teleport this state from Charlie to Alice. 
\Cref{fig:threeResults} presents the fidelity of state teleportation for each of the six cardinal logical Pauli eigenstates (yields from post-selection are also shown).   
We also demonstrate teleportation of logical S and T states. The logical T states are prepared non-fault-tolerantly \cite{Li_2015, Lodyga_2015,Brown2020universal, Singh2022, Lao2022}.
We use a fault-tolerant measurement of logical Y (see Supplemental Material S1.2) to perform logical tomography on the teleported state.

Next, we demonstrate a protocol to create a logical Bell pair between Alice and Charlie without any direct coupling of their logical qubits; results are presented in \cref{fig:threeResults}.  This is achieved through teleportation, and verified by estimating  $\langle$XX$\rangle$ and $\langle$ZZ$\rangle$ across those distant qubits to obtain state fidelity bounds on the teleported Bell state.  Post-selecting on the correct preparation of the initial states (yield 1.2\%), we observed a logical Bell-state fidelity bounded between 85.4\% and 91.4\%.

Finally, we demonstrate logical gate teleportation, which is an enabling primitive in most fault-tolerant architectures for quantum computing.  In gate teleportation, a logical state $|\psi\rangle$ on C is teleported to the state $U|\psi\rangle$ on A for some logical gate $U$. Conversely the logical state can be teleported from A to C; see Supplemental Material S5 where full details of the protocol are provided.
\Cref{fig:threeResults} presents the fidelities associated with teleporting a range of initial states through a selection of logic gates, from A to C and from C to A.  

We use state tomography to measure the fidelity of the output state; we do not perform logical process tomography here. 
We note that the lattice surgery between A and B is of lower fidelity than the lattice surgery between B and C (qubit 55 is to blame), impacting the reported fidelity of the C to A protocol.

\paragraph{Fast resets to counter leakage.}
\label{sec:reset}

Errors caused by leakage out of the qubit subspace occur at low rates on superconducting qubits in IBM quantum processors.  However, in QEC experiments involving many repeated rounds of syndrome extraction, while stochastic Pauli errors and measurement errors are being identified and corrected, leakage left uncorrected can eventually compound to become the dominant logical error.  This effect has been observed in stability experiments across a number of recent IBM quantum processors, where the expected exponential decrease in the probability of error plateaus after a number of syndrome measurement rounds~\cite{harper2025characterising, drouet2026defects, mclauchlan2026}.  This effect has been attributed to uncorrected leakage errors.

A key innovation in the revision 2 Nighthawk quantum processor \device is the ability to perform a fast hardware reset --- a new capability for IBM processors.  Reset gadgets offer several advantages; they give shorter syndrome extraction circuits for a given fault distance and, for our purposes, the capability enables us to address leakage.  Qubits that are reset during the repeated syndrome extraction measurements are returned to the qubit subspace, thereby eliminating the potential for leakage errors to compound on these qubits.

Here we demonstrate leakage suppression using fast resets in stability experiments; see \Cref{fig:stability}. Our choice of stability experiment involves a $4\times4$-sized stability patch. We consider two locations, and perform stability experiments for each location both with, and without, using fast resets on the ancilla qubits at each measurement round.

In these standard stability experiments, as with the implementations of QEC we consider in this work, only the ancilla qubits are measured in each round of syndrome extraction. The fast reset will counter the accumulation of leakage errors on these ancilla qubits, but leakage on data qubits is left unaddressed because these data qubits are not measured.  To explore the full potential of resets as a method to eliminate leakage errors, we perform a `walking' stability experiment~\cite{McEwen2023}. 
The `walking circuit' exchanges roles between data and ancilla qubits in each location every round by means of a teleportation step in the stabilizer readout circuits. This means that every qubit involved in the circuit is measured and reset every two rounds.

The results of our stability experiments are shown in \Cref{fig:stability}.  Without resets, we see a tapering in the reduction of probability of error in the stability experiment as the number of syndrome measurement rounds is increased. With resets in the standard stability experiment (only ancilla qubits are reset) the probability of error decreases more rapidly but also tapers. 
With the walking circuit stability experiment, where every qubit is reset every two rounds, the rate of reduction in the probability of error is the most significant.  
At 170 rounds we are in the regime of $10^{-7}$ probability of error.  For this data point, with 50 million runs we observed 8 failed decoding attempts. 
Even at the remarkable speed of \device, we are at the limit or our ability to probe further~\footnote{To put this in perspective, 50 million runs of a 170 round stability experiment of this size comprises 603.5 billion two qubit gates and 170.8 billion total measurements including mid-circuit measurements.}.

For the logical qubit experiments reported in this work, fast combined measurement and reset are used in all syndrome extraction circuits. We do not use walking circuits.

\paragraph{Future directions.}

We have used a commercially accessible device to demonstrate high-fidelity entangling operations between logical qubits whose logical error per round is well below~$1\%$. These results made essential use of defect exclusion strategies. We have also shown that stabilizer readout circuits designed to reduce leakage can achieve very low logical error rates for error-corrected logic by means of a stability experiment. Our demonstrations can be regarded as a prototype towards a large-scale quantum computer capable of performing algorithms with logical error rates at $10^{-8}$ or below. However, we must progress further before we can produce a quantum computer capable of solving industry-relevant problems.

First, we require a larger number of logical qubits performing with much better fidelity than those we have demonstrated here. This will ultimately require quantum processors with many more qubits and with more connectivity. Continued improvement in the components of these quantum processors will decrease logical error rates, but to achieve orders-of-magnitude improvement we need larger arrays of qubits that enable us to produce bigger codes to suppress the probability of logical error exponentially in the code distance. In the future, we might expect that superconducting-qubit devices will be supplemented by long-range couplers; perhaps to connect qubit arrays in a modular fashion, or otherwise to realise more sophisticated quantum low-density parity-check codes that can encode more logical qubits at a higher distance with fewer physical qubits~\cite{yoder2025tourgrossmodularquantum}. It will be valuable to continue to develop our benchmarking tools to assess the performance of these highly connected devices as experiments progress.

In this work, we have limited our attention to logical Clifford operations. However, a quantum computer will require logical non-Clifford operations to complete a universal gate set. To complement experimental demonstrations of these types of logic gates, there is a need for new benchmarking tools as well as better simulation methods to investigate the noise that occurs throughout these operations that take us beyond our more conventional noise sampling methods. These tools will be essential as we continue to build a universal error-corrected logic gate set for a industry-scale quantum computer.

\begin{acknowledgements}
We are grateful for conversations with P. Bonderson, M.~Carroll, A.~Cross, B.~Het\'{e}nyi, A.~Javadi, R.~Mishmash, E.~Pritchett, M.~Shaw, V.~Siddhu, S.~Singh, K.~Siva, M.~Takita and T.Yoder.  BJB is grateful for the hospitality of the Niels Bohr Institute at the University of Copenhagen. We acknowledge support from the Intelligence Advanced Research Projects Activity (IARPA), under the Entangled Logical Qubits program through Cooperative Agreement Number W911NF-23-2-0223 (R.H., B.J.B., and S.D.B.).  The views and conclusions contained in this document are those of the authors and should not be interpreted as representing the official policies, either expressed or implied, of IARPA, the Army Research Office, or the U.S. Government. The U.S. Government is authorized to reproduce and distribute reprints for Government purposes notwithstanding any copyright notation herein. 
\end{acknowledgements}

\section{S1 - Memory Experiments}

Here we discuss in more detail the memory experiments discussed in the main text. \Cref{fig:threshold} shows the location of the distance-5 patch and the operation of the code with qubits 55, 62 and 95 removed from the patch. 

\subsection{S1.1 - Logical error per round in memory experiments}

\begin{figure}[p]
\centering
\begin{tikzpicture}[every node/.style={inner sep=0}]
  \node[anchor=north west] (a) at (-3,0)
    {\includegraphics[width=.33\linewidth]{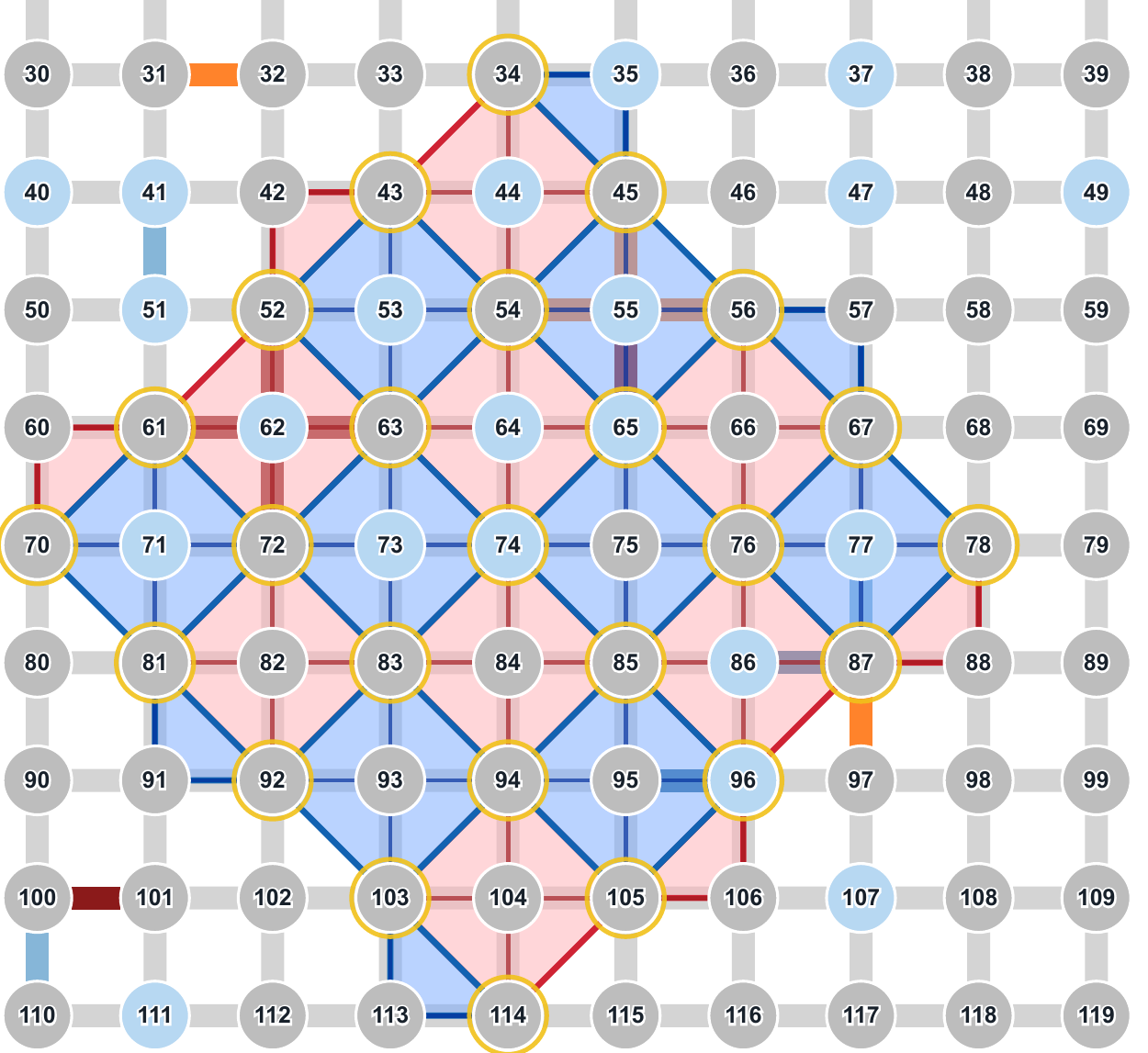}};
  \node[anchor=north west] (b) at ([xshift=2mm]a.north east)
    {\includegraphics[width=.33\linewidth]{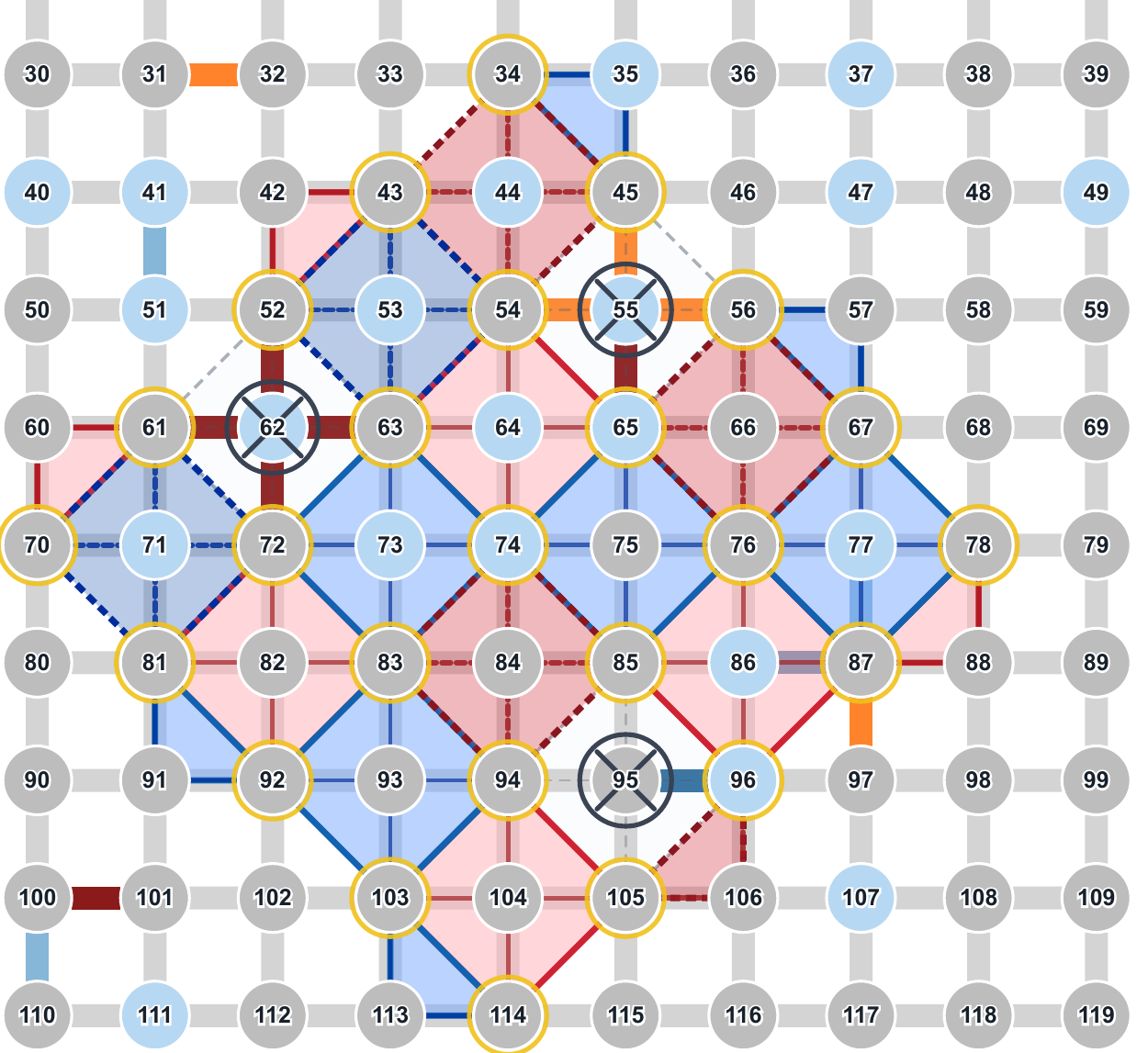}};
  \node[anchor=north west] (c) at ([xshift=2mm]b.north east)
    {\includegraphics[width=.33\linewidth]{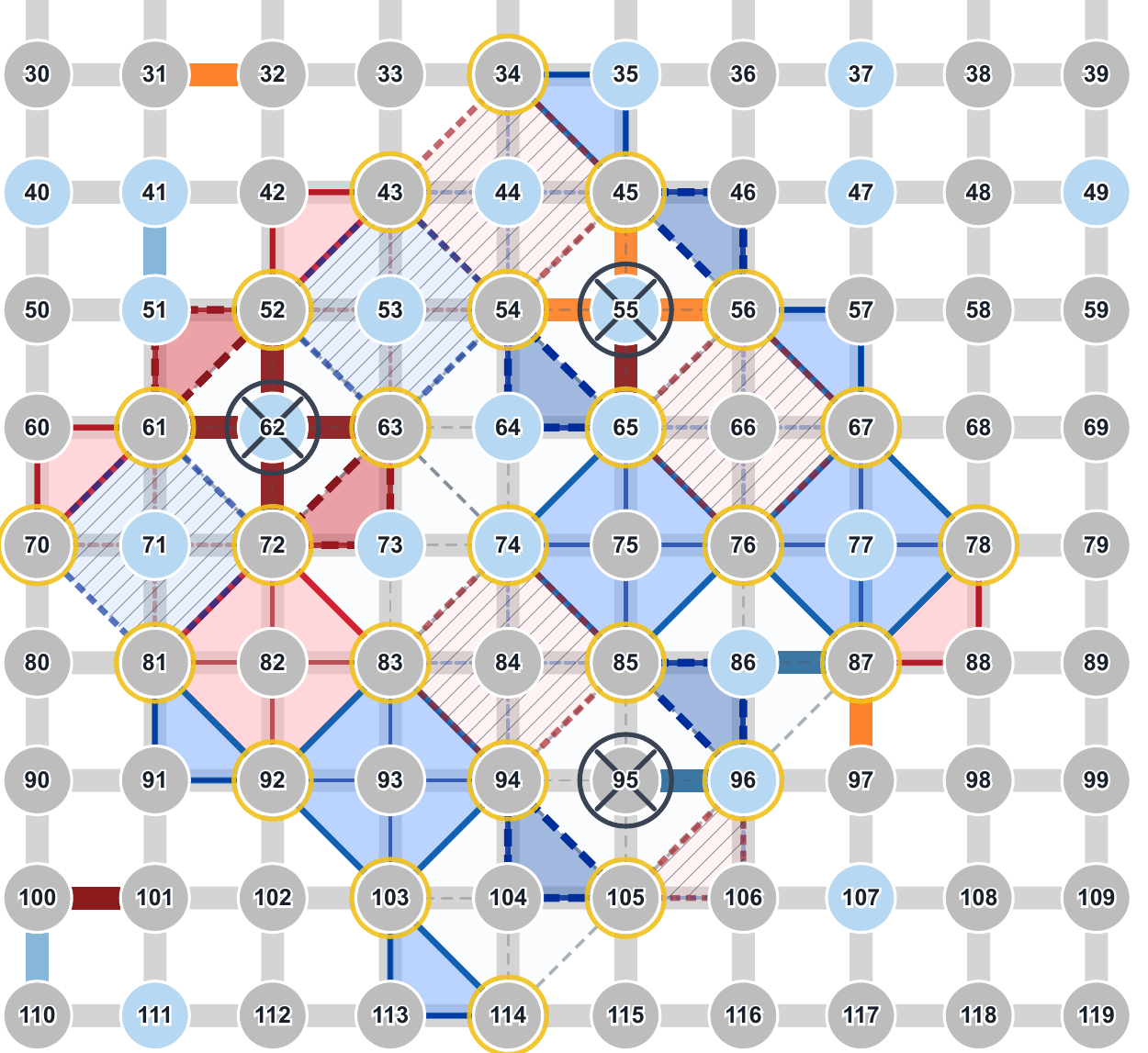}};
  \node[anchor=north west] at (-3,0.5)
    {\includegraphics[width=1\linewidth]{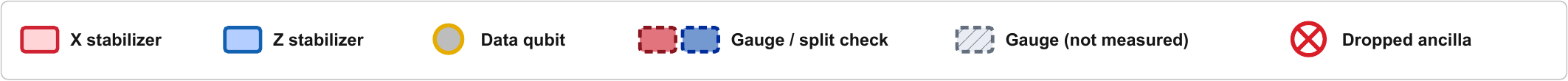}};

  \node[anchor=north] at ([yshift=-1mm]a.south) {(a)};
  \node[anchor=north] at ([yshift=-1mm]b.south) {(b)};
  \node[anchor=north] at ([yshift=-1mm]c.south) {(c)};

\end{tikzpicture}
\caption{\textbf{Memory experiments with ancilla re-purposing on a distance-5 code.}  (a) Placement of a standard distance-5 rotated surface code on the device. Here we overlay the code on a calibration of the map at the time the experiments were conducted. The fidelity bands on the gates and the qubits are designed to highlight less performant couplers and qubits. As can be seen, the patch includes two poorly performing locations, the couplers surrounding qubit 62 and qubit 55, and a poorly performing coupler connecting qubits 95 and 96. (b) This shows the first round of the ancilla re-purposing strategy designed to remove qubits 62, 55 and 96 from the code. Several stabilizers become gauge checks because of the way stabilizers are measured in the other round. Namely the stabilizers measured on ancillas 71 and 73, those on 44 and 66 and those on 84 and 106. In this round the stabilizers that would be measured by the removed ancillas are not measured. (c) Here we re-purpose neighbouring ancillas to measure the missing stabilizers, as two two-body stabilizers. These anti-commute with some of the stabilizers (pale shaded stabilizers), however as described previously the shaded patches can be used in the form of gauge checks (forming 8-body/6-body detectors) which commute with the combined two-body stabilizers.}
\label{fig:threshold}
\par\vspace{8pt}


\centering
\includegraphics[width=.9\linewidth]{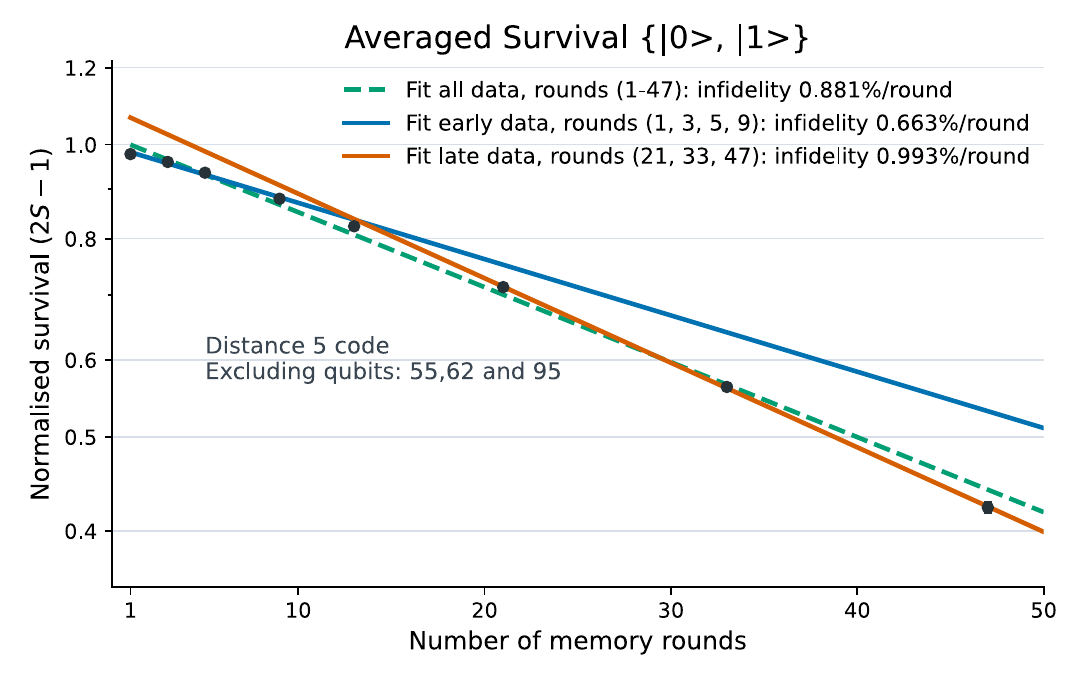}
\caption{\textbf{Fitting logical error per round in memory experiments.}  Comparison of different fitting approaches for logical error per round in memory experiments, based on different choices in the number of rounds. The data points are the averaged non-excited and excited state survival chances (here, $|0\rangle$ and $|1\rangle$)  for 50,000 shots per circuit per round count per basis. The average survival percentage at a particular round is plotted, normalised as $2S-1$ (where $S$ is the survival probability). This allows them to be plotted on a log axis. The fit is to $Ap^m+0.5$, where $A$ is a SPAM factor and $p$ is the decay factor. Single logical qubit fidelity is calculated as $(p+1)/2$. The error rate per round is $1-\mathrm{fidelity}$. The 0.5 factor comes from averaging the excited and non-excited state given that this is a trace preserving experiment \cite{Harper2018}. We note that this is identical to the fitting procedure used in Ref.~\cite{sundaresan2022}. }
\label{fig:fitting}
\end{figure}

For all memory experiments, to obtain logical error per round we fit over 50 syndrome extraction rounds (47 memory rounds after the first detectors are fully formed) using the process described in the caption of \cref{fig:fitting}. As noted in Ref.~\cite{kan2026demonstrationlogicalerrorsuppression}, fitting the logical error per round to only the first few rounds of error correction tends to yield smaller values. A likely explanation for this effect is the accumulation of leakage errors on the data qubits. 

As such leakage represents real and significant error processes that should be captured in a reported logical error per round, all of our memory fits have been over a large range of rounds rather than just the first few rounds. Despite the slight, non-exponential behavior observed, we believe this gives a more accurate depiction of the suppression of errors in the quantum memory. \Cref{fig:fitting} provides an illustration of this effect.

\subsection{S1.2 - Logical measurement and state preparation in the Y basis}
We implement both logical Y basis state preparation and logical Y basis measurement for the surface code using circuits discussed in Ref.~\cite{Gidney_2024_ybasis}, for both distance-3 and -5 surface codes. With the appropriate padding of syndrome extraction circuits (padding of 2 for distance-3 codes, and 3 for distance-5), we achieve fault-tolerant Y preparation and readout in 3 and 4 rounds, respectively, meaning that the appropriate distance is maintained.  See Ref.~\cite{Gidney_2024_ybasis} for details.

\begin{figure}
    \centering
    \includegraphics[width=0.6\linewidth]{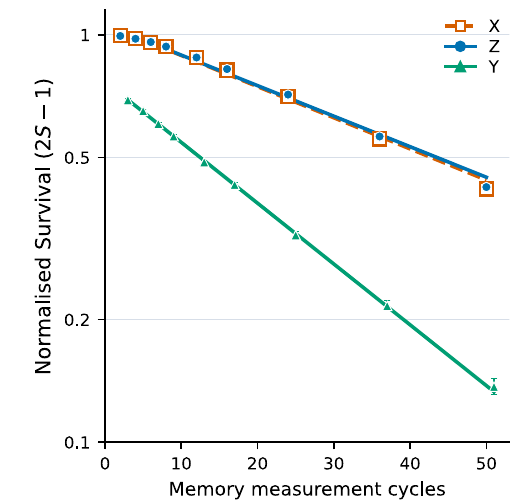}
    \begin{tabular}{@{}l@{\hspace{2em}}r@{}}
    \toprule
    \textbf{Basis} & \textbf{LER} \\
    \midrule
    $X$ & 0.872\% \\
    $Z$ & 0.855\% \\
    $Y$ & 1.690\% \\
    \bottomrule
    \end{tabular}
    \caption{\textbf{Y basis memory experiments.}  Memory experiments performed in the logical X, Y, and Z bases using the distance-5 code at location q34 as in Appendix~A.  Ancilla re-purposing is used in the syndrome extraction circuits as described in that Appendix, but not in the logical Y state preparation or measurement circuits. The reduced y-axis intercept for the logical Y basis experiment indicates the additional logical SPAM associated with the logical Y preparation and measurement circuits. The  logical error rate is calculated as discussed in the caption of \cref{fig:fitting} and the per round rate is shown in the table below the graph. Note that the logical error rate per round of the Y state is a close match to the sum of the other two, as predicted in Ref.~\cite{Gidney_2024_ybasis}.  }
    \label{fig:placeholder}
\end{figure}

We test these circuits with memory experiments, preparing the logical qubit in a Y basis state and performing a final measurement in the logical Y basis after repeated QEC rounds.  
The logical error rate per round for each basis is presented in \cref{fig:placeholder}, comparing with the same memory experiments in the X and Z bases.  Ancilla re-purposing was not used for the logical Y basis preparation and readout, however it was used for the syndrome extraction circuits in the memory rounds. The errors in the Y-state preparation and readout are logical state preparation and measurement (SPAM) errors. Because we are fitting to a decay curve these SPAM errors only affect the y-axis intercept not the reported logical error rate. However as can be seen in \cref{fig:placeholder} the SPAM errors for preparing/reading the Y-state are considerably greater than for the X/Z state.  

It has been shown in simulation that the Y-basis error rate for the surface code is approximately the sum of the Z and X error rates~\cite{Gidney_2024_ybasis}. Our results are consistent with this observation, matching an independent flip model $p(y)\approx p(x) +p(z) -2p(x)p(z)$  to a high degree of accuracy.  

\section{S2 - Ancilla re-purposing}

\begin{figure}
\centering
\begin{tikzpicture}[
    inner sep=5pt,
    outer sep=0pt
]

\node[
    draw=black,
    rounded corners=8pt,
    line width=0.8pt] (round0) at (-4.8,0) {
    \includegraphics[width=0.18\textwidth]{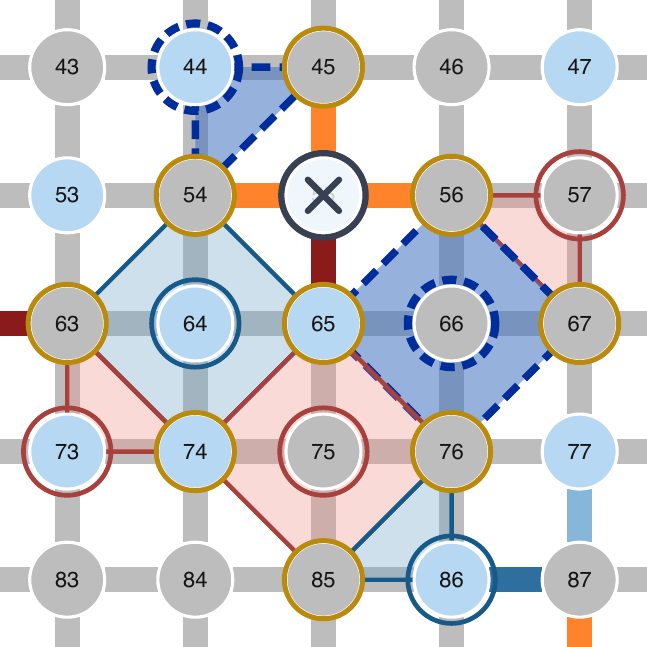}
};

\node[
    draw=black,
    rounded corners=8pt,
    line width=0.8pt,
    right=0.2cm of round0] (round1)  at (-2.9,0) {
    \includegraphics[width=0.18\textwidth]{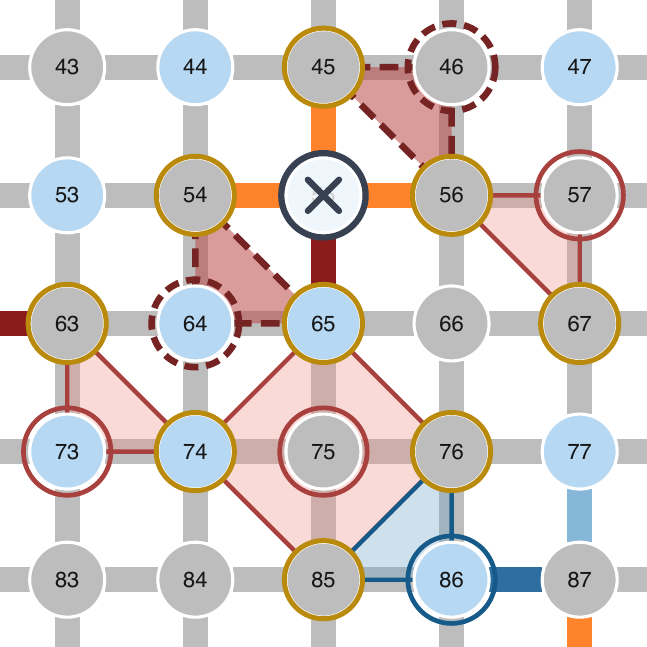}
};

\node[
    draw=none,
    inner sep=0pt,
    right=0.2cm of round1] (legend) at (0.7,-1.2) {
    \includegraphics[width=0.12\textwidth]{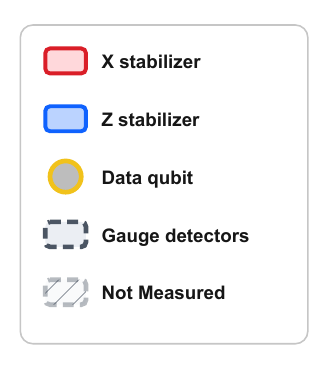}
};
\node[
    draw=none,
    inner sep=0pt,
    right=0.2cm of round1] (legend) at (0.7,1.5) {
    \includegraphics[width=0.12\textwidth]{figures/UniversalLegend.pdf}
};

\end{tikzpicture}
\caption{\textbf{Ancilla re-purposing.}  Ancilla re-purposing \cite{Leroux2025snakes} is illustrated using a distance-3 code, with top data qubit 45 and two-body stabilizer to the left used to identify this code location as q45(L).  For the purposes of this illustration we remove qubit 55 and the couplers attached to it. \textbf{(Left)} Round 1.  The four-body stabilizer around the excised qubit is not measured. Shading on the stabilizers measured using ancilla qubits 44 and 66 indicates they are gauge operators, as they anti-commute with two-body operators we will measure in the next round. \textbf{(Right)} Round 2.  The missing stabilizer will be inferred by measuring two, two-body gauge operators with ancilla qubits 64 and 46.  These anti-commute with the gauge operators measured on qubits 44 and 66 in the previous round, but commute with their product. Note as shown here on the second round we do not measure the stabilizers on 44 and 66 in this round. Where one of these two stabilizers is an edge stabilizer (as in this example) we can alter the syndrome extraction round cycles to allow these to be measured at the same time as the new two-body stabilizers (without causing adverse hook errors). This means they are only gauge stabilizers in one round and normal stabilizers in the other. Somewhat surprisingly we observed no benefit in doing this.}
\label{fig:snl}
\end{figure}
\begin{figure}
\centering
\begin{minipage}[t]{1\linewidth}
  \vspace{0pt}
  \centering
  \includegraphics[width=0.65\linewidth]{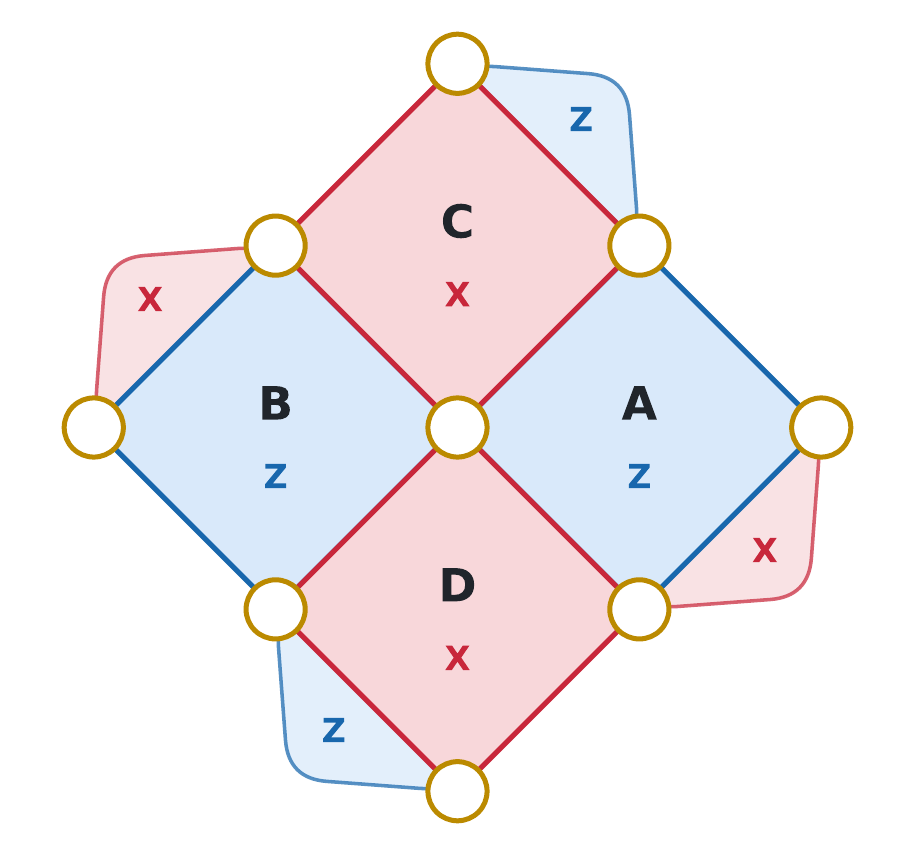}
\end{minipage}\hfill
\begin{minipage}[t]{1\linewidth}
  \vspace{0pt}
  \centering
  \small
  \setlength{\tabcolsep}{4pt}
  \begin{tabular}{lrrrrr}
    \toprule
    Location & None & A & B & C & D \\
    \midrule
    q34(R) & 1.46 & \textbf{0.72} & 3.50 & 1.83 & 1.28 \\
    q45(L) & 1.42 & 1.95 & 1.34 & \textbf{0.67} & 1.73 \\
    q56(R) & 0.99 & 1.50 & 1.21 & 1.68 & \textbf{0.93} \\
    q43(L) & 2.18 & 4.73 & \textbf{1.25} & 5.47 & 3.17 \\
    q54(R) & 2.55 & \textbf{1.94} & 2.43 & 2.33 & 2.66 \\
    q54(L) & 0.99 & \textbf{0.80} & 1.35 & 1.40 & 0.80 \\
    q65(L) & 1.43 & 1.13 & 1.08 & 1.50 & \textbf{0.93} \\
    q52(R) & 1.75 & 1.43 & 1.77 & \textbf{0.88} & 2.11 \\
    q63(L) & 1.20 & 1.37 & 1.34 & 1.41 & \textbf{1.19} \\
    q74(R) & 1.14 & \textbf{0.76} & 1.12 & 1.08 & 1.10 \\
    \bottomrule
  \end{tabular}
\end{minipage}
\caption{\textbf{Comparing ancilla re-purposing strategies.}  Effect on logical error per round of re-purposing different ancillas for the distance-3 codes that sit in the footprint of the distance-5 code presented in \cref{fig:threshold}. The diagram at the top indicates the ancilla that has been removed. The code location on the device is identified by the data qubit at the top of the diagram. The direction indicator indicates where the two-body stabilizer is in relation to the indicator qubit; in the case of the diagram, this is an (R) layout. The table presents the logical error per round where no ancilla is removed, and the logical error per round where each ancilla is removed individually using ancilla re-purposing. The lowest logical error per round for each location is highlighted in bold. In all instances, there is always at least one ancilla for which its removal improves performance.}
\label{fig:phoenix_d3_snl_comparison}
\end{figure}

Variability in the performance of individual components on IBM quantum processors is a significant limiting factor in the resulting performance of logical qubits.  In particular, some qubits are sufficiently under-performant that strategies to exclude their use are required.  In the experiments reported here, we use the strategy of Ref.~\cite{Leroux2025snakes} where we re-purpose ancilla qubits so as to avoid underperforming qubits and couplers.

\Cref{fig:snl} shows an example of how ancilla re-purposing proceeds, using a distance-3 code.  Unlike the usual syndrome extraction circuit for the surface code, which measures all commuting stabilizers in a single round, the ancilla-re-purposing circuit takes two rounds to measure all stabilizers and gauge operators, with some detectors spanning both rounds.
Where it is an ancilla that is being excluded, the distance of the code can often be maintained, although this may not be possible if there are too many defects close to each other.

The benefit of ancilla re-purposing is very dependent on the quality of the components (qubits and couplers) being excluded as well as the quality of the ones that are being drafted in to replace them. 
We tested each possible four-body replacement for the distance-3 codes in the code-patch used for our distance-5 experiment. This gave us the original 10 codes plus 40 variations to test. As shown in \cref{fig:phoenix_d3_snl_comparison}, ancilla re-purposing does not offer an advantage in every case. However, somewhat surprisingly, for every location tested there was at least one ancilla that could be excluded to improve the error rate.

\section{S3 - Noise}

\subsection{S3.1 - Crosstalk}

\begin{figure}
    \centering
    \begin{minipage}[t]{1\linewidth}
  \vspace{0pt}
  \centering

    \includegraphics[width=0.95\linewidth]{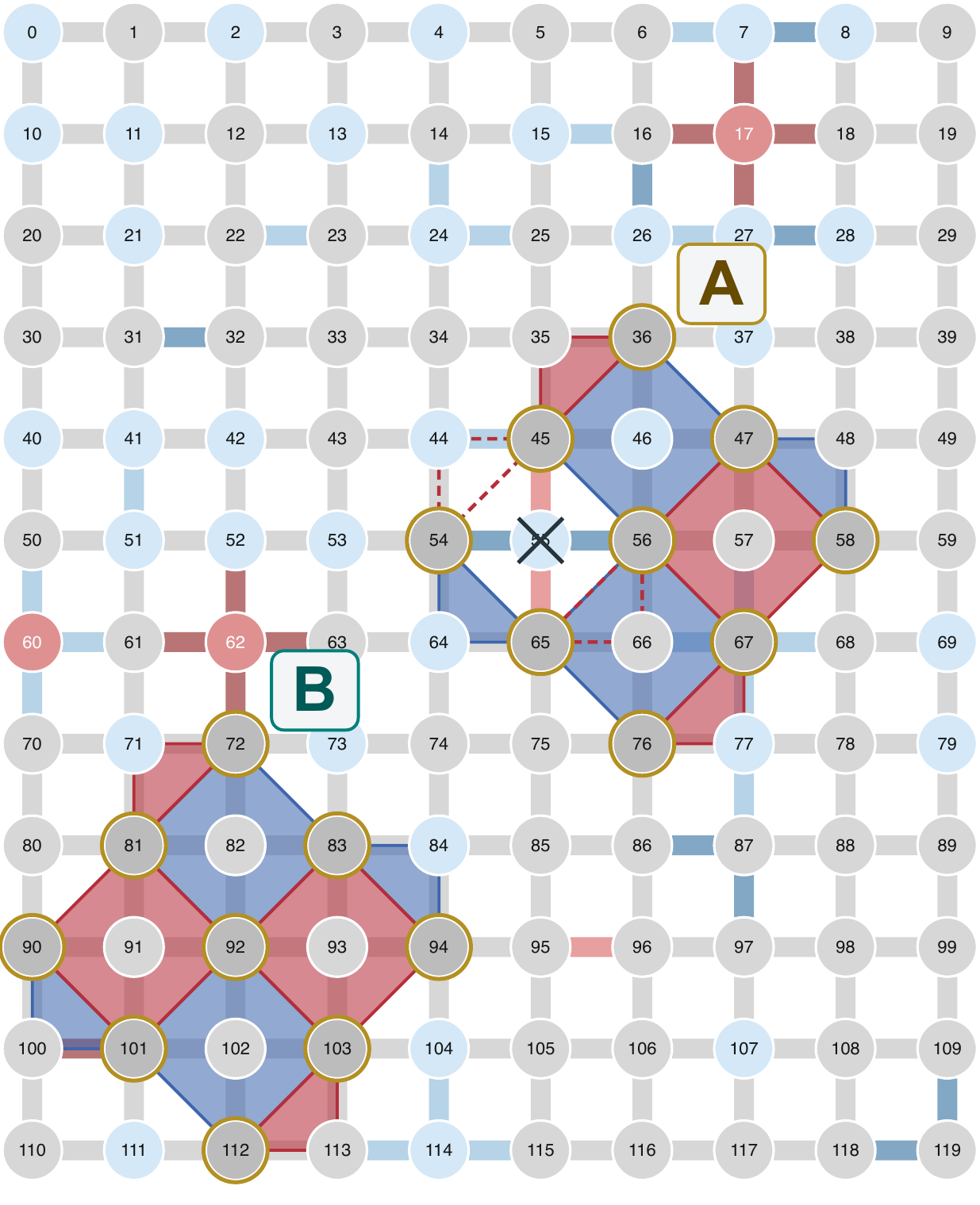}
    \end{minipage}
    \hfill
\begin{minipage}[t]{1\linewidth}
  \vspace{0pt}
  \centering
  \small
\setlength{\tabcolsep}{4pt}
\begin{tabular}{@{}cccccc@{}}
\toprule
\multicolumn{2}{c}{\textbf{Input}} &
\multicolumn{2}{c}{\textbf{Per-round error}} &
\textbf{Correlation} & \textbf{Lift} \\
\cmidrule(lr){1-2}
\cmidrule(lr){3-4}
\textbf{A} & \textbf{B} & \textbf{A} & \textbf{B} & & \\
\midrule
$|0\rangle$ & $|0\rangle$ &
\shortstack{1.91\%{\scriptsize (+0.19)}} &
\shortstack{0.86\%{\scriptsize (+0.03)}} & 0.0152 & 2.16 \\
$|0\rangle$ & $|+\rangle$ &
\shortstack{1.95\%{\scriptsize (+0.23)}} &
\shortstack{0.83\%{\scriptsize (+0.03)}} & 0.0167 & 2.29 \\
$|+\rangle$ & $|0\rangle$ &
\shortstack{1.03\%{\scriptsize (+0.08)}} &
\shortstack{0.86\%{\scriptsize (+0.03)}} & 0.0141 & 2.48 \\
$|+\rangle$ & $|+\rangle$ &
\shortstack{1.02\%{\scriptsize (+0.07)}} &
\shortstack{0.82\%{\scriptsize (+0.01)}} & 0.0073 & 1.79 \\
\bottomrule
\end{tabular}
\end{minipage}
    \caption{\textbf{Crosstalk between logical qubits.}  \textbf{(Top.)} Placement of two distance-3 logical qubits operated simultaneously, but independently. Logical qubit A is being operated with qubit 55 removed, using the techniques described in Appendix~B. \textbf{(Bottom.)} The per-round error is the logical error per round when the logical qubits are operated simultaneously. The number in brackets is the increase in the logical error rate (in percentage points) compared with operating the logical qubits individually. Lift is the ratio of the probability that one logical qubit has an error given an error on the other, to its unconditional error probability.}
    \label{fig:crosstalk}
    \label{tbl:crosstalk}
\end{figure}

Crosstalk is a concern for many solid-state quantum processors, due to its potential impact on QEC performance.  For recent (revision 2 and 3) IBM Heron quantum processors, crosstalk has been very minimal~\cite{harper2025characterising}. In \device, we identify higher levels of crosstalk arising from the simultaneous operation of various two-qubit couplers. As shown in \cref{tbl:crosstalk} this crosstalk has a small but significant impact on error correction. \Cref{fig:crosstalk} shows the location of two independently operated logical qubits (qubit A is being operated with ancilla re-purposing to exclude qubit 55 as discussed in section Appendix B).  These memory experiments are operated both individually and then simultaneously and decoded independently.  When they are operated simultaneously we form a joint probability distribution for each QEC round. Using a Walsh-Hadamard transformation, we transform these probability distributions into channel eigenvalues which are then fit to exponential decay curves. The fit values are transformed back to a SPAM free per-round joint probability distribution \cite{Harper2019b}. This joint distribution is analyzed in \cref{tbl:crosstalk}, revealing the correlations between the two logical qubits caused by crosstalk.   
The \emph{lift} indicates that the probability that B fails given that A fails is roughly twice B’s marginal failure probability (and \textit{vice versa}).

\subsection{S3.2 - Noise Characterisation}
We used averaged circuit eigenvalue sampling (ACES)~\cite{flammia2021b,hockings2024,Hockings_QuantumACES_jl_design_noise_2025} to provide an initial detailed characterization for the syndrome extraction circuits of a distance-5 code and of the two logical qubit Bell-state experiment.  
This characterization data is used to construct a detector error model that can inform decoding.

For circuits not directly characterized by ACES, we construct an ACES-informed stim noise model as follows. ACES splits an error correcting code into layers of simultaneously operating gates, and characterizes the noise on each gate in each layer. If ACES has been used to characterize a particular code/position on a particular day we can use this data directly. For different times we can use changes in the calibration data supplied for the days in question to scale the ACES data appropriately. For a code position or type that is not identical to the code which ACES has characterized, each target layer of the error correction code is matched to the characterized layers of the same type by looking for the largest operation overlap. We call this the relevant layer. The noise characterization of individual gates in the relevant layer is reused (scaled by calibration shifts) when every physical operation that occurs in the layer can be matched to an operation in the characterized ACES layer. Otherwise, we take the channels for the same gate type from the relevant layer, scale each one for the target operation using calibration data, and average their Pauli-error probabilities component by component. If the relevant layer has no channels of that gate type then we average over other ACES layers that do have the noise estimate for that gate type.

The resulting model is used to generate the detector error model which is used throughout in conjunction with a BeliefMatching decoder \cite{higgott2023improved}. The only exception is the stability experiment where PyMatching \cite{higgott2021pymatching} was used, due to the very large size of the decoding problem.

\section{S4 - Compressed lattice surgery}

\begin{figure}
\centering
\begin{tikzpicture}
  \node[anchor=south west, inner sep=0] (img) at (0,0) {%
    \includegraphics[width=0.95\linewidth]{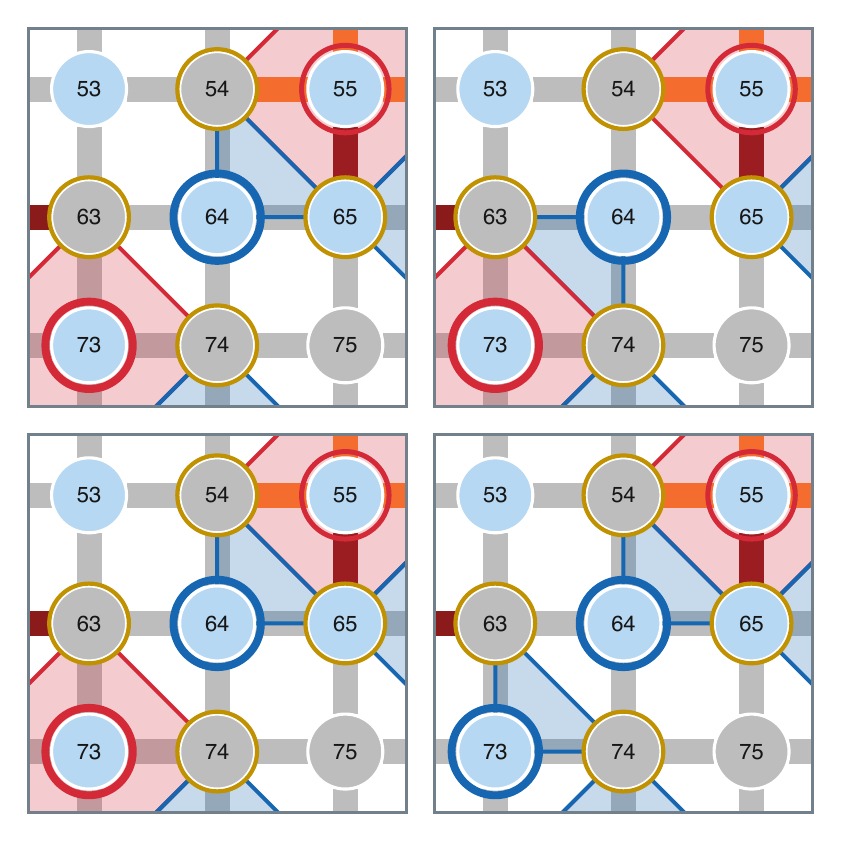}%
  };
  \begin{scope}[x={(img.south east)}, y={(img.north west)}]
    \tikzset{panel label/.style={
      anchor=north west,
      fill=white,
      draw=black!55,
      rounded corners=2pt,
      inner xsep=4pt,
      inner ysep=2pt,
      font=\bfseries\large
    }}
    \node[panel label] at (0.045,0.962) {A};
    \node[panel label] at (0.530,0.962) {B};
    \node[panel label] at (0.045,0.480) {C};
    \node[panel label] at (0.530,0.480) {D};
  \end{scope}
\end{tikzpicture}
    \caption{\textbf{Compressed lattice surgery using ancilla re-purposing.}  Two methods to place independent surface codes next to each other are shown. Here we focus on qubit 64, which is an ancilla qubit that would normally service both codes. Panels A and B demonstrate ancilla sharing where the two logical qubits have access to the ancilla on different rounds. While the ancilla is reset between rounds, this approach nonetheless provides a channel for errors to spread between codes. In panels C and D, one logical qubit (here the top one) has exclusive access to the ancilla. The other logical qubit uses qubit 73 to alternately measure the four-body X stabilizer on qubits 63, 74, 72 and 83 (panel C) and the two-body Z stabilizer between qubits 63 and 74 (panel D) in different rounds.  The two codes never share ancilla qubits in this approach.}
    \label{fig:shared_stabilizer}
\end{figure}

\begin{figure}
    \centering
    \includegraphics[width=1\linewidth]{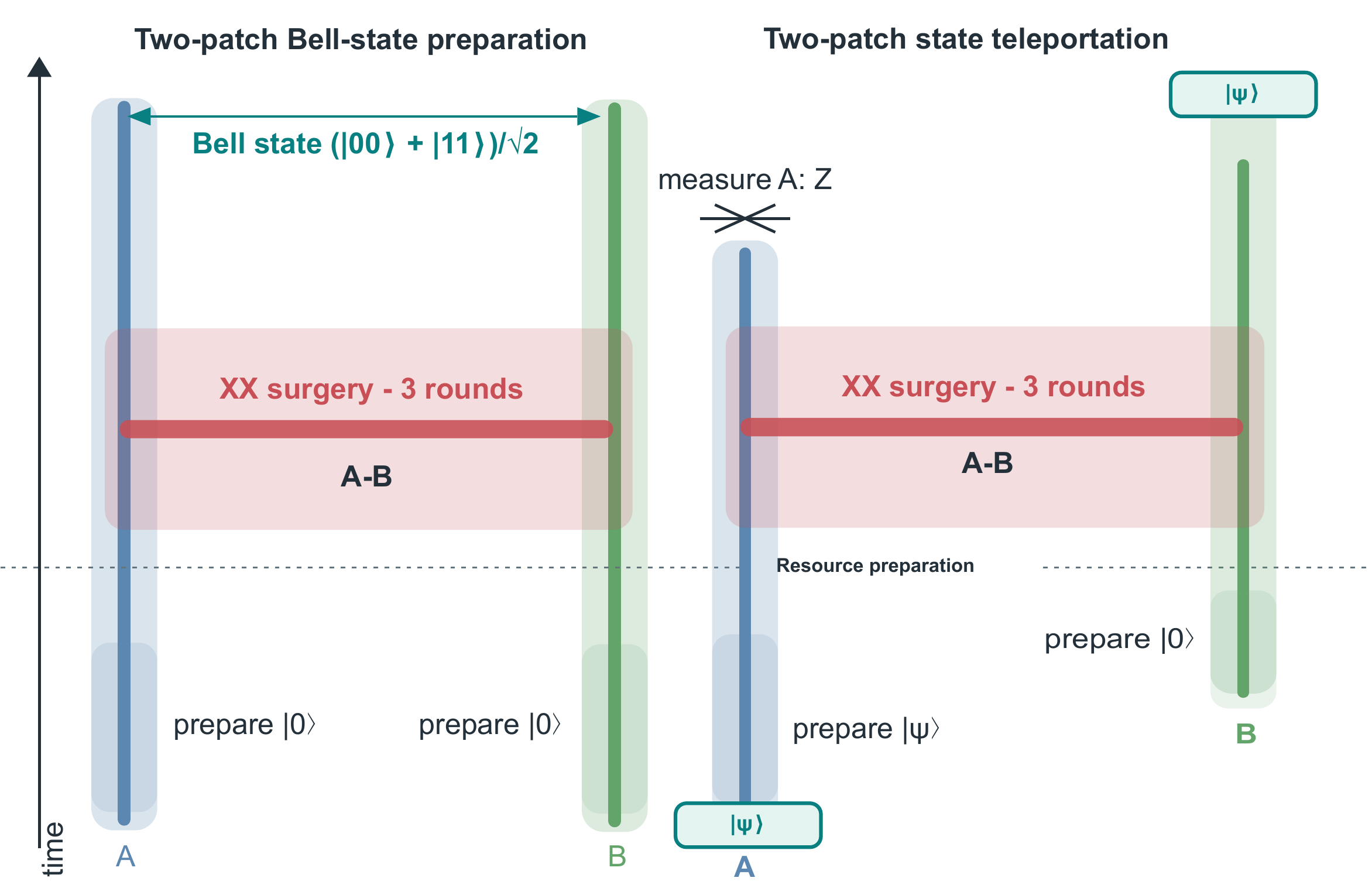}
    \caption{\textbf{Pipe diagrams of two-qubit lattice experiments showing the flow of information.} }
    \label{fig:pipes}
\end{figure}

\begin{figure*}
    \centering
    \begin{tikzpicture}
    \tikzset{panel label/.style={
      anchor=north west,
      fill=white,
      draw=black!55,
      rounded corners=2pt,
      inner xsep=4pt,
      inner ysep=2pt,
      font=\bfseries\large
    }}
  \node[anchor=south west, inner sep=0] (img) at (0,0) {%
    \includegraphics[width=0.95\linewidth]{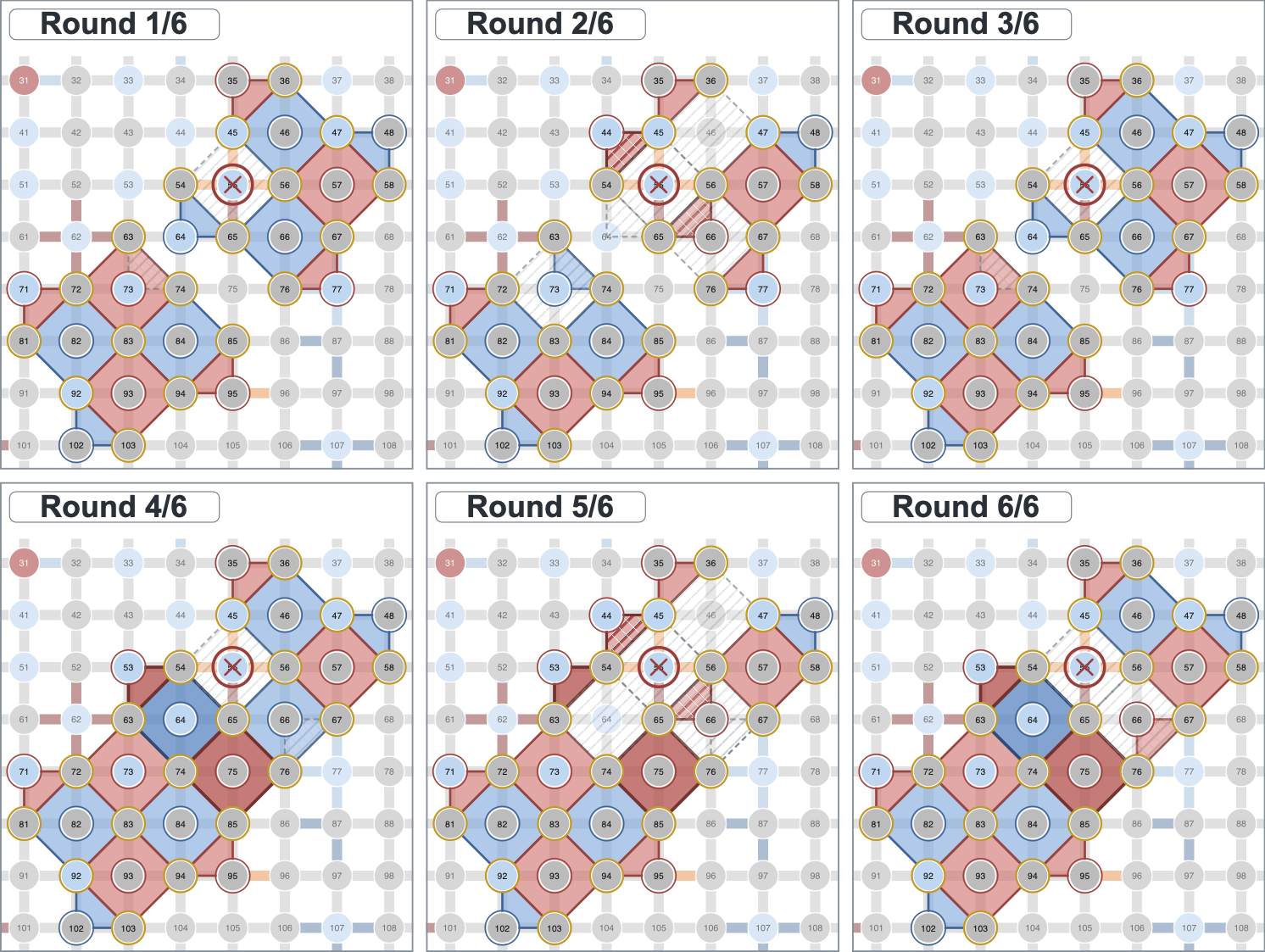}};
    \node[panel label] at (3.1,12.7) {Alice (A)};
    \node[panel label] at (3,7.1) {Bob (B)};
    \end{tikzpicture}
    \caption{\textbf{Circuits used to perform lattice surgery on two logical qubits.} This is a six round experiment. In rounds 1 to 3 we are preparing the logical qubits. We deal with the shared ancilla (qubit 64) by allowing Alice to use it. Bob uses round 1 to prepare the X (red) stabilizer on qubit 73. In logical qubit A we do not want to use qubit 55, so we do not use it this round. In round 2 Bob uses qubit 73 to measure the Z (blue) stabilizer on 63 and 74. This commutes with all his other stabilizers. Alice measures the X (red) stabilizer on qubit 55 by splitting the four-body stabilizer into two two-body gauge checks measured on qubit 44 and qubit 66. She does not measure the four-body checks on qubits 46 and 66 or the two-body check on qubit 64 in this round. 
    However, the checks on qubits 64 and 46 that Alice will measure going forward now become gauge checks and from now on must be multiplied together to ensure they commute with the two gauge checks on qubits 44 and 66.
    Round 3 is the same as round 1, at this point we have detectors formed for all of the stabilizers (treating the two-body gauge checks as one stabilizer and the gauge checks on qubits 64 and 46 as another stabilizer). Round 4 is the first round of lattice surgery. The newly measured stabilizers on 53 and 75 are marked in darker red. These are the ones that will give us the outcome of the `XX' measurement. They must be read three times to maintain distance. The gauge checks measured on qubit 64 and qubit 46 need to be combined to form a stabilizer. Note we do not read the boundary stabilizer on qubit 77 this round. We are going to stop using qubit 77. In round 5 we measure the stabilizer that would be read by qubit 55 as two two-body gauge operators on qubit 44 and the re-purposed qubit 66. We do not read out qubit 64 or 46. We do not use qubit 77. In the final round we re-purpose qubit 66 to read out the two-body stabilizer that would normally be read by qubit 77. This two-body stabilizer commutes with all the others. At this point we can destructively read all the qubits and determine the appropriate observables. The entire circuit is a distance-3 circuit. }
    
    \label{fig:c_overview}
\end{figure*}

As discussed in the main manuscript, we use ancilla re-purposing techniques to run distance-3 logical qubits in close proximity. \Cref{fig:shared_stabilizer} shows the two methods that we use to make this possible. While we normally use the methods shown in panels C and D of that figure, sometimes other mitigation techniques make it necessary to use the techniques shown in A and B. We note this when we present different circuits. Compressed lattice surgery proceeds as illustrated in \cref{fig:c_overview}, together with our ancilla re-purposing strategy for the two logical qubit experiments. The first three QEC rounds are used to prepare the qubits in the correct logical state (three rounds enable all detectors to be formed), followed by lattice surgery and readout. A distance of 3 is maintained throughout. In \cref{fig:pipes} we show pipe diagrams for the flow of information in the Bell-state experiment and the teleport experiment.

More specifically the exact sequences used in each protocol are as follows.  

\textbf{Bell-state fidelity.} We prepare each logical qubit in the code space by initializing the qubits in  $|0\rangle$ and performing three syndrome extraction rounds. With the ancilla re-purposing we require three rounds to fully form all the detectors in the code. 
The codes are then combined by measuring the stabilizers along the seam. 
We perform three syndrome extraction rounds of lattice surgery.
We then measure out the data qubits (in the $Z$ or $X$ basis) and infer the appropriate logical observable. 
If we measure out in the $Z$ basis, the observable is the product $\bar{Z}_1\bar{Z}_2$ obtained on readout. 
If we measure out in the $X$ basis, the observable is $\bar{X}_1\bar{X}_2$ from readout, combined with the outcome $m_X$ of that same logical operator obtained during lattice surgery. 
The Bell state we have projected into is determined by the product of the central $X$-type merge-check outcomes on the seam.

\textbf{Teleportation protocol.} 
We initially prepare the source in one of the four cardinal states \(\{\ket{0}_L,\ket{1}_L,\ket{+}_L,\ket{-}_L\}\), and prepare the destination in the $\ket{0}_L$ state.
As discussed above we perform, in total, three rounds of syndrome extraction (one to project into the code space and two more to form all the detectors).
The source is then coupled to the destination  through lattice surgery, by measuring the stabilizers across the seam.
After three lattice surgery rounds, the data qubits belonging to the original source region are destructively measured in the $Z$ basis. Finally the destination logical qubit is measured in the same basis in which the source was prepared.

\section{S5 - Three logical qubit experiments}

\begin{figure*}
    \centering
     \begin{tikzpicture}
    \tikzset{panel label/.style={
      anchor=north west,
      fill=white,
      draw=black!55,
      rounded corners=2pt,
      inner xsep=4pt,
      inner ysep=2pt,
      font=\bfseries\large
    }}

    \node[anchor=south west, inner sep=0] (img) at (-6,4) {%
        \includegraphics[width=0.33\linewidth]{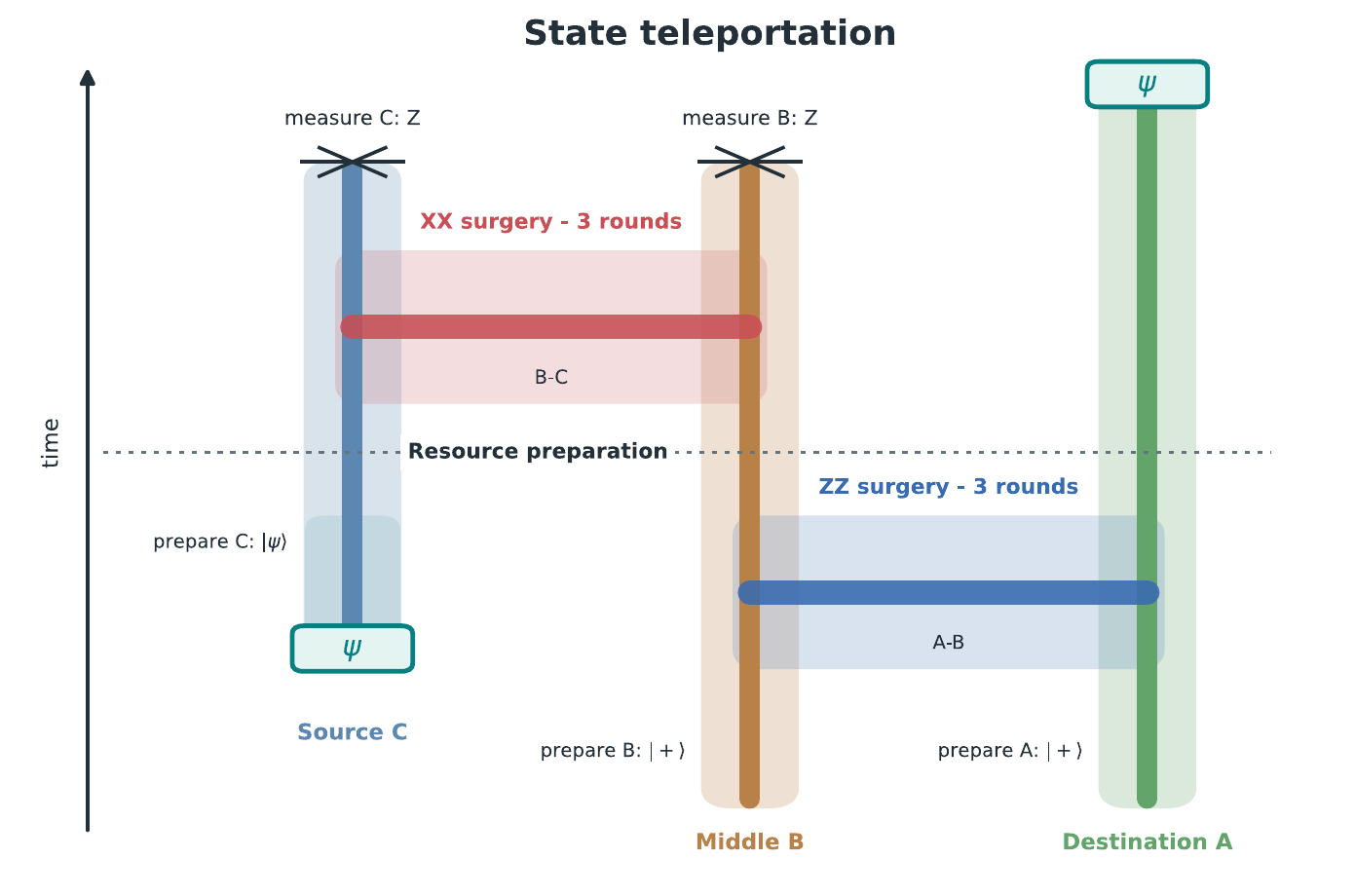}};
    \node[anchor=south west, inner sep=0] (img) at (0,4) {%
      \includegraphics[width=0.33\linewidth]{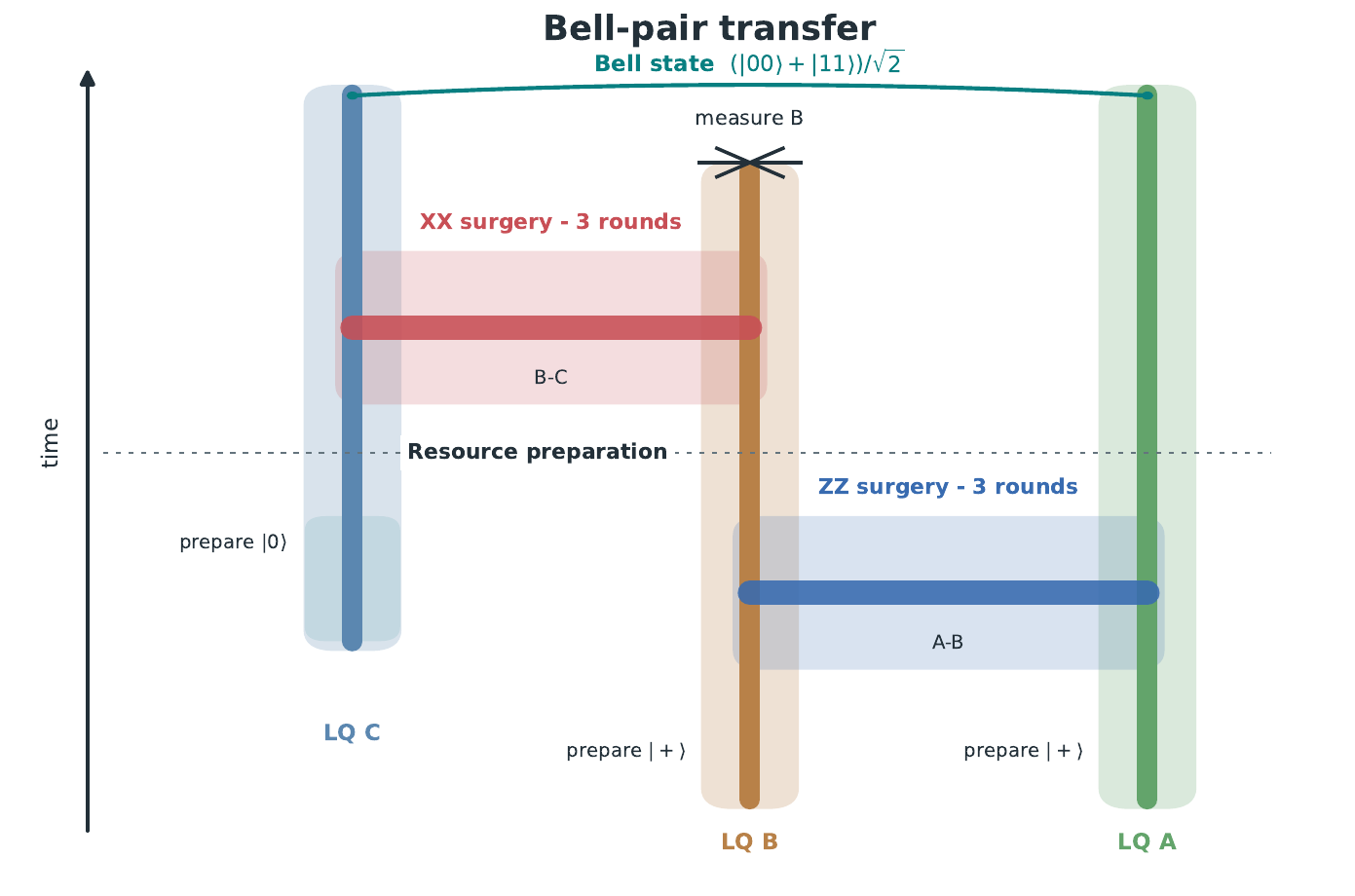}};
    \node[anchor=south west, inner sep=0] (img) at (6,4) {%
      \includegraphics[width=0.33\linewidth]{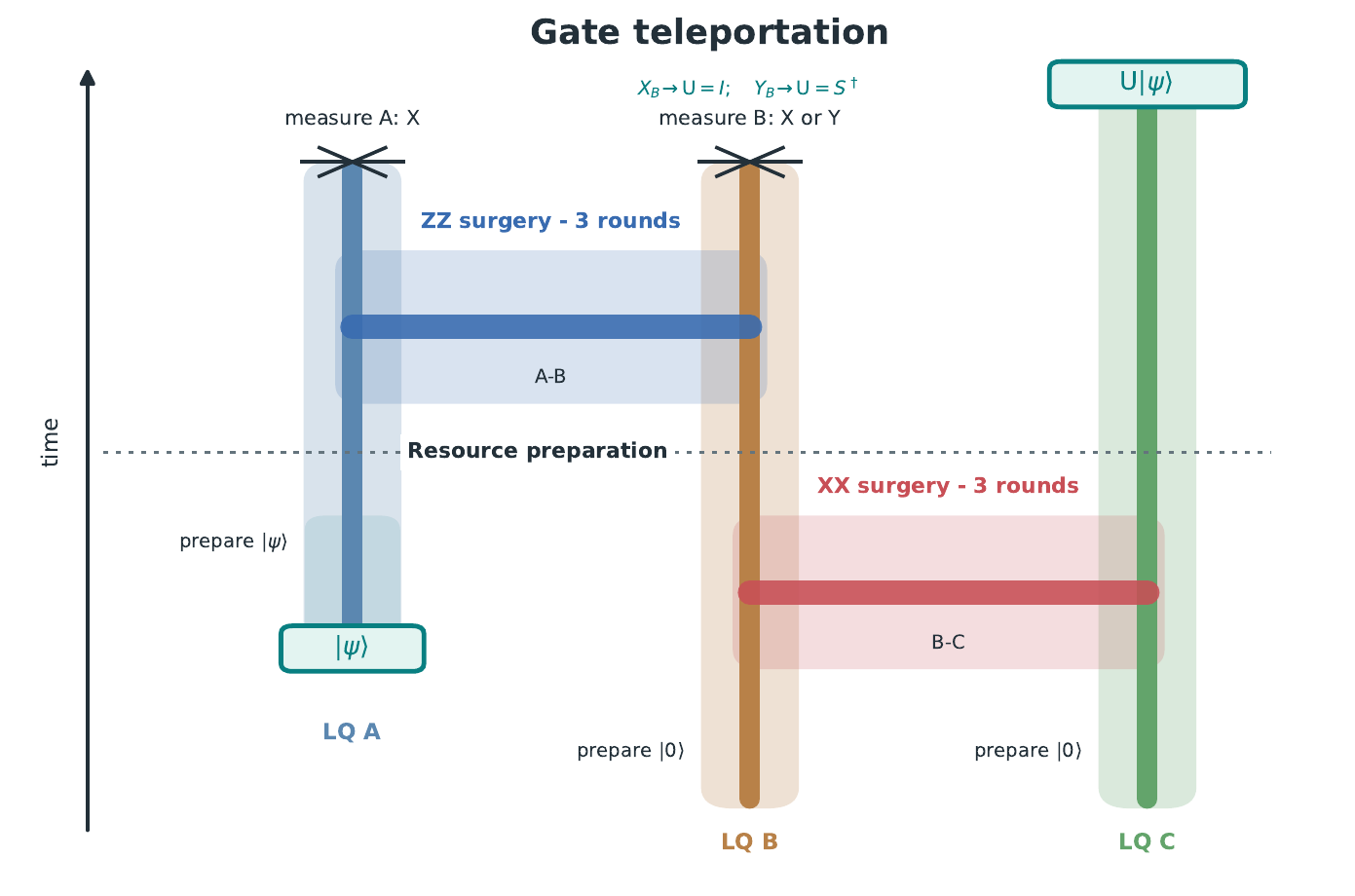}};
    
    \node[anchor=south west, inner sep=0] (img) at (-3,-15) {%
    \includegraphics[width=0.7\linewidth]{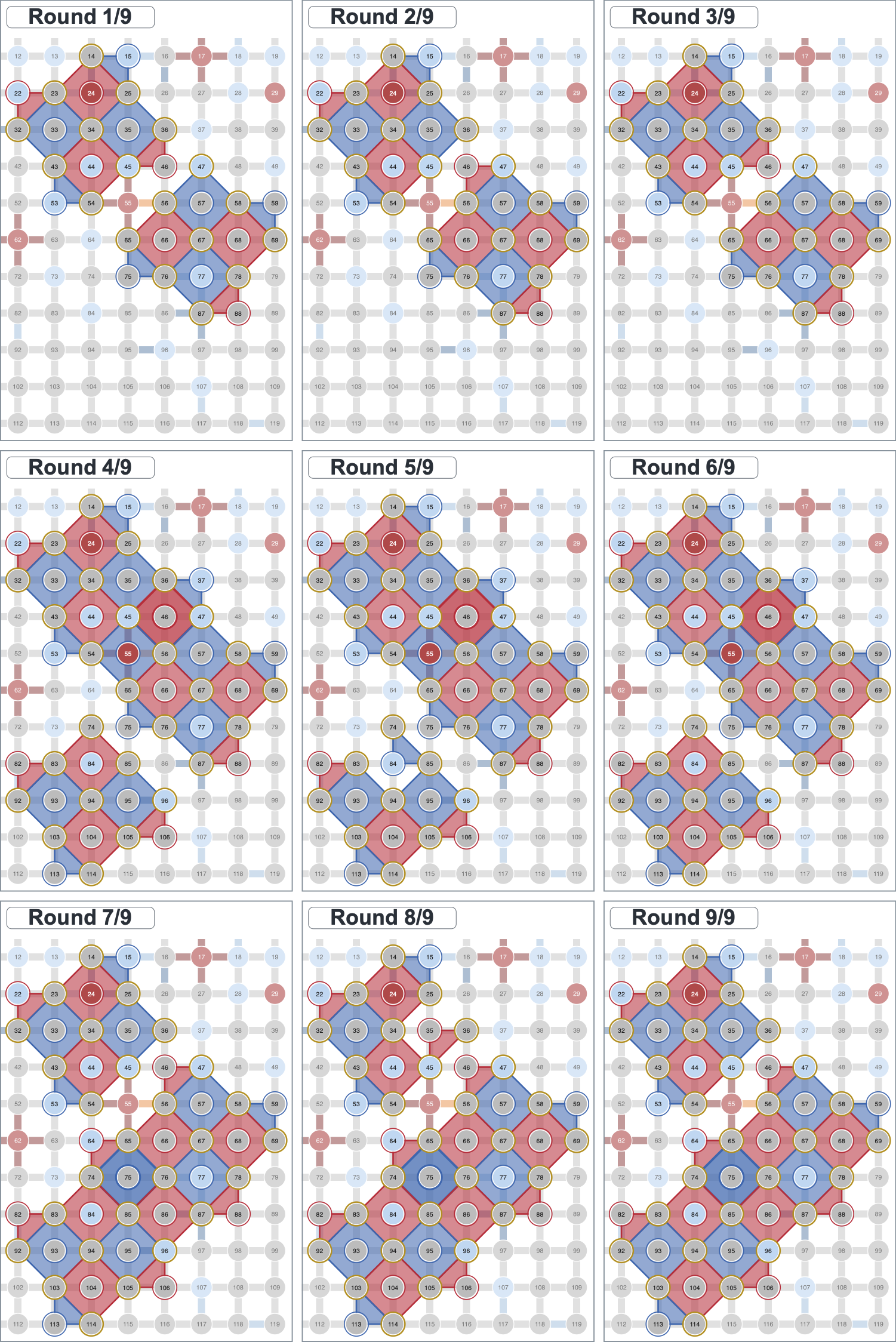}};
    \node[panel label] at (-5.2,2) {Alice (A)};
    \node[panel label] at (-3.5,0) {Bob (B)};
    \node[panel label] at (-5.5,-7) {Charlie (C)};

    \end{tikzpicture}
    \caption{\textbf{Three logical qubit schedule for Alice, Bob, and Charlie.} (\textbf{top}) Pipe diagrams of thre-qubit lattice experiments. \textbf{(bottom)} Schedule. In rounds 1 to 3, Alice and Bob prepare their logical qubits. There they use the ancilla preparation schedule described in panels A and B of \cref{fig:shared_stabilizer}. In rounds 4 to 6 Charlie enters and prepares her logical qubit while Alice and Bob entangle theirs. Charlie uses the strategy described in panels C and D of \cref{fig:shared_stabilizer} to allow Bob to use qubit 75. Finally in rounds 7 to 9 Charlie and Bob entangle their qubits, while Alice reverts to the strategy described in panels C and D of \cref{fig:shared_stabilizer} to avoid using qubit 46.}
    \label{fig:pipes3}
    \label{fig:threeway}
\end{figure*}

We extend the two-patch experiments to three adjacent distance-three logical qubits, Alice, Bob and Charlie  labeled A, B and C. 
We can fit these three logical qubits on the device through the use of compressed lattice surgery techniques previously discussed. 
The middle patch B participates in two successive lattice-surgery measurements: one with A and one with C. 
We illustrate the flow of information with pipe diagrams in \cref{fig:pipes3}.
\Cref{fig:threeway} also shows the location of the logical qubits and details the ancilla re-purposing used here.

We detail the protocols below, but generally they follow the same sequence of events. 
The first two of the logical qubits to be entangled are prepared in the code-space.
Preparation of the patch not involved in the first round of surgery runs concurrently with that first lattice surgery.
The second lattice surgery occurs while the logical qubit that is not involved is maintained through syndrome extraction circuits. 
Finally the logical qubits are destructively read-out.

In total each circuit consists of nine main syndrome-extraction rounds: three for initial preparation,
followed by three rounds for each lattice-surgery measurement other than when we need logical-$Y$ readout on logical qubit B. We can only commence the three-round fault-tolerant $Y$ readout after we have finished the second set of lattice surgery. In that case we require an additional three syndrome measurement rounds to maintain distance-3 circuits.

The same arrangement supports three related experiments. 
We transfer a prepared logical state from C to A, transfer one half of an A--B Bell pair to C, and transfer a state from A to C while selecting a logical operation through the measurement basis of B. 
In each case the measurement outcomes from the seams of the two lattice surgeries determine the appropriate Pauli frame which is used in analysing the final logical observable.

The sequences for the three protocols are described separately below.

\textbf{State teleportation (C to A).}
We first prepare logical qubits A and B in $\ket{+}_L$, using three syndrome-extraction rounds to establish their code states and detectors.
We then perform three rounds of $\bar{Z}_A\bar{Z}_B$ lattice surgery to prepare a logical Bell pair. 
During these rounds, the input state is prepared on logical qubit C. 
Next, we perform three rounds of $\bar{X}_B\bar{X}_C$ lattice surgery while A continues in memory.
B and C are destructively measured in the $Z$ basis, completing the
logical Bell measurement. 
The outcomes of the two lattice-surgery
measurements, together with the destructive readout, determine the
Pauli frame of the state transferred to A. 
We finally measure A in the basis appropriate to the input state. 

\textbf{Bell-pair teleportation (A--B to A--C).}
The first six rounds are the same as in state teleportation: A and B
are encoded and projected into a Bell state by
$\bar{Z}_A\bar{Z}_B$ lattice surgery, while C is prepared in
$\ket{0}_L$. 
We then perform three rounds of
$\bar{X}_B\bar{X}_C$ lattice surgery and destructively measure B in
the $Z$ basis. 
This transfers B's half of the Bell pair to C, up to
a Pauli frame determined by the measurement outcomes. 
To characterize
the resulting A--C pair, we destructively read out A and C together
in either the $Z$ or $X$ basis. 
The  $\bar{Z}_A\bar{Z}_C$ and $\bar{X}_A\bar{X}_C$ expectations provide
bounds on the Bell-state fidelity \cite{mclauchlan2026}.

\textbf{Gate teleportation (A to C).}
For the gate-teleportation experiment we reverse the order of the two lattice-surgery operations. 
B and C are prepared in $\ket{0}_L$, then coupled by three rounds of $\bar{X}_B\bar{X}_C$ lattice surgery.
The input state is prepared on A during this operation. 
We subsequently perform three rounds of $\bar{Z}_A\bar{Z}_B$ lattice surgery and
destructively measure A in the $X$ basis. 
The measurement basis chosen for B determines the operation on the state transferred to C: an
$X$-basis readout gives the identity-teleportation reference, while a $Y$-basis readout gives an $S^\dagger$ operation. 
The measurement outcomes determine the Pauli-frame correction. 
Finally, C is read out in $X$, $Y$, or $Z$ to tomographically reconstruct the output state.

    

\bibliography{Hex} 

\end{document}